\documentclass[%
 reprint,
 amsmath,amssymb,
 aps,
]{revtex4-2}

\usepackage{graphicx}
\usepackage{dcolumn}
\usepackage{bm}
\usepackage{color}
\usepackage[table]{xcolor}
\usepackage{array}
\usepackage{booktabs}
\usepackage{mathrsfs,bm}
\usepackage{multirow}
\usepackage{hyperref}

\newcommand{\aap}{{A\&A}}
\newcommand{\apjs}{{Astrophys.~J.~Supp.}}

\newcommand{\apjl}{{Astrophysics. Journal. Letters.}}

\def\lsim{\;\rlap{\lower 2.5pt
    \hbox{$\sim$}}\raise 1.5pt\hbox{$<$}\;}
\def\gsim{\;\rlap{\lower 2.5pt
    \hbox{$\sim$}}\raise 1.5pt\hbox{$>$}\;}

\begin{document}

\preprint{APS/123-QED}

\title{Minkowski Functionals of Large-Scale Structure as a Probe of Modified Gravity } 

\author{Aoxiang Jiang}
\email{jax9709@mail.ustc.edu.cn}
\affiliation{CAS Key Laboratory for Research in Galaxies and Cosmology, Department of Astronomy, University of Science and Technology of China, Hefei, Anhui, 230026, P.R.China}
\affiliation{School of Astronomy and Space Sciences, University of Science and Technology of China, Hefei, Anhui, 230026, P.R.China}
\author{Wei Liu}
\affiliation{CAS Key Laboratory for Research in Galaxies and Cosmology, Department of Astronomy, University of Science and Technology of China, Hefei, Anhui, 230026, P.R.China}
\affiliation{School of Astronomy and Space Sciences, University of Science and Technology of China, Hefei, Anhui, 230026, P.R.China}
\author{Baojiu Li}
\affiliation{Institute for Computational Cosmology, Department of Physics, Durham University, South Road, Durham DH1 3LE, UK}
\author{Cristian Barrera-Hinojosa}
\affiliation{Instituto de F\'isica y Astronom\'ia, Universidad de Valpara\'iso, Gran Breta\~na 1111, Valpara\'iso, Chile}
\affiliation{Department of Computer Science, Durham University,  South
Road, Durham DH1 3LE, UK}
\affiliation{Institute for Computational Cosmology, Department of Physics, Durham University, South Road, Durham DH1 3LE, UK}
\author{Yufei Zhang}
\affiliation{College of Mathematics and Physics, Leshan Normal University, Leshan 614000,China}
\author{Wenjuan Fang}
\email{wjfang@ustc.edu.cn, corresponding author}
\affiliation{CAS Key Laboratory for Research in Galaxies and Cosmology, Department of Astronomy, University of Science and Technology of China, Hefei, Anhui, 230026, P.R.China}
\affiliation{School of Astronomy and Space Sciences, University of Science and Technology of China, Hefei, Anhui, 230026, P.R.China}

\date{\today}

\begin{abstract}
{
{
In this study, we explore the potential of utilizing the four 
Minkowski functionals, which can fully describe the morphological
properties of the large-scale structures, as a robust tool for 
investigating the modified gravity, particularly on non-linear 
and quasi-linear scales. 
With the assistance of the N-body simulation, we employ the
Minkowski functionals to probe the Hu-Sawicki $f(R)$ gravity model.
The focus is on understanding the morphorlogical properties extracted 
by the Minkowski functionals and their sensitivity to modified gravity.
Our analysis involves a comprehensive examination of the cosmic 
variance arising from finite simulation volumes. By systematically 
varying smoothing scales and redshifts, we quantify the information 
encoded in the Minkowski functionals measured from the dark-matter 
density field. The goal is to assess the capacity of the Minkowksi
functionals to constrain the model and explore potential improvements 
through their combination.
Additionally, we investigate the impact of using biased tracers 
such as dark matter halos and the halo occupation distribution 
galaxies on the modified gravity signatures within the Minkowksi 
functionals of the LSS. Furthermore, we evaluate the influence of 
the redshift space distortion on the observed results.
In summary, our study suggests that the Minkowski functionals of the 
large-scale structures hold promise as a stringent tool for 
constraining modified gravity and offer valuable insights into 
the morphological features of the cosmic web.
}
}

\end{abstract}

\maketitle


\section{Introduction}
\label{sec:intro}

Einstein's theory of general relativity (GR) 
has achieved great success within the last hundred years 
due to its remarkable agreement with various experimental 
and observational tests, ranging from tests in the 
laboratory and the Solar System to the emission of 
gravitational waves from merging binary objects \cite{2016PhRvL.116f1102A,2017ApJ...848L..13A},
among others. However, the aforementioned tests probe 
the law of gravity on small scales, and any attempt to 
extend GR's applicable scale to the whole Universe 
could represent a challenge. The discovery of the late-time 
accelerated Hubble expansion in the 1990s \cite{1998AJ....116.1009R,1999ApJ...517..565P}, 
which since then has been supported by several 
observational evidences (e.g., \citep{2014ApJ...782...74H, 2020A&A...641A...6P, 2021arXiv210513543A, 2018MNRAS.477.1639Z} for recent works), 
raises the possibility that GR might be flawed on cosmological scales.

As alternatives to GR, theories of modified gravity (MG) 
challenge the applicability of GR on large scales (see e.g., \cite{2010LRR....13....3D,2012PhR...513....1C,2017PhR...692....1N} for MG reviews) 
arguing that the need to introduce additional dark energy, 
which has negative pressure and does not have any non-gravitational support, follows from the inappropriate application of GR on large scales. By modifying the way gravity works, these theories offer the possibility of cosmic self-acceleration without introducing an additional dark energy component. 

One way to test MG theories is by using the large-scale 
structure (LSS) of the Universe. 
Since the structure of the cosmic density field is 
formed from tiny primordial density fluctuations that 
grow large and non-linear due to gravitational collapse, 
in general the predictions of MG gravity do not match 
those from GR. For instance, in $f(R)$-gravity~\cite{2004PhRvD..70d3528C, 2005PhRvD..71f3513C, 2007PhRvD..76f4004H}, 
which is a popular class of MG theories that 
contains a nonlinear function $f(R)$ of the Ricci 
scalar $R$ in the Einstein-Hilbert action, 
the enhancement of gravity leads to more non-linear growth.
{That indicates to us the great importance to 
focus on the small scales where the traditional 2-point 
statistics cannot provide us with full information of 
the structure and higher-order statistics should be considered.}

Thanks to a wealth of galaxy redshift surveys such as SDSS \cite{sdss2000}, BOSS \cite{2017MNRAS.470.2617A}, and WiggleZ \cite{2010MNRAS.401.1429D}, one can extract the evolution information encoded in the 3D galaxy distribution by 2-pt statistics such as the galaxy power spectrum, or correlation function, and thus probe any departure from GR and put constraints on the parameter space of MG models. Furthermore, the ongoing and upcoming redshift surveys such as DESI \cite{2016arXiv161100036D}, CSST \cite{2019ApJ...883..203G}, and Euclid \cite{2018LRR....21....2A}, will also provide us with huge survey volume and a large number of galaxies to obtain more stringent constraints from the LSS.

The 2-pt statistics mentioned above can fully capture 
the information contained in the LSS if it were Gaussian. 
However, extracting the non-Gaussian information efficiently 
is a more complicated task, especially with the increasing 
number of galaxies in future surveys. N($>$2)-pt statistics 
can offer us additional information, but require much more 
computational resources and are challenging to be modeled accurately. 
As an alternative, the Minkowski functionals (MFs), which can fully describe the morphological properties of the LSS, have been proposed \cite{1994A&A...288..697M}. Compared with N-pt statistics, MFs can principally capture information at all orders and they are easier to measure. Since introduced in 1990s \cite{1994A&A...288..697M}, they are widely used to study primordial non-Gaussianity \cite{2006ApJ...653...11H, 2012MNRAS.423.3209P}, weak lensing \cite{2012PhRvD..85j3513K, 2021MNRAS.507.1421M, 2012ApJ...760...45S}, neutrino mass \cite{2020PhRvD.101f3515L, WeiLiu2021, liu2023probing}, and redshift space distortions \cite{2013MNRAS.435..531C, PhysRevD.105.023527, 2022PhRvD.105j3028J}, among others.   

MFs of the two-dimensional weak lensing maps have been shown to have the potential to detect the signatures of MG~\cite{2015PhRvD..92f4024L,2017MNRAS.466.2402S}. MFs of the three-dimensional LSS have also been proposed as a probe of gravity by \cite{2017PhRvL.118r1301F}. Specifically, they measured the MFs of the matter distribution from N-body simulations, and found strong signals in the MFs for discriminating the normal-branch Dvali-Gabadadze-Porrati (nDGP) \cite{2000PhLB..485..208D} and $f(R)$ \cite{2007PhRvD..76f4004H} MG models from GR. However, to apply this new probe in real galaxy surveys, one needs to take into account systematic effects such as redshift distortion and tracer bias. Moreover, since the modification to gravity is usually scale- and redshift-dependent, it is worthwhile to study how the constraining power of the MFs varies with these two parameters.

In this work, motivated by the unprecedented precision achievable by upcoming LSS surveys, we study several aspects regarding the application of the LSS's MFs as a probe of modified gravity in galaxy surveys.
First, based on N-body simulations, we construct the Fisher matrix to quantize the constraining power of the MFs on modified gravity parameters and study how sample variance or survey volume affects the constraints by measuring the MFs in simulation boxes with different sizes. Then we analyze the effects of MG on the MFs measured from the matter density field and see how they change with the smoothing scale and redshift. We also investigate how the constraints on modified gravity parameters change with these two quantities, and examine the improvement in the combined constraint of different smoothing scales and redshifts. Tracer bias is also taken into account in this work. By comparing the difference in the MFs between MG and GR measured from the distributions of halo and matter respectively, we discover the significant impact of halo bias on the MG signature in the MFs. (For a detailed analysis of how redshift distortion affects the MFs of LSS, see our recent work of \cite{2022PhRvD.105j3028J}.) We then study how the constraints on modified gravity parameters change when using biased tracers of the LSS.

The investigation presented in this work contributes to deepening our understanding on MG, especially its dependence on scale, redshift, and tracer bias. They also set an example for the applications of the MFs of LSS in constraining cosmological parameters other than the MG parameters, such as, the fractional matter density $\Omega_m$, neutrino mass $M_\nu$, and the amplitude of the density fluctuations $\sigma_8$. The layout of this paper is as the follows. We briefly describe the Hu-Sawicki f(R) theory and the simulations used in this work in Sec.~\ref{sec:mod}. In Sec.~\ref{sec:method}, we introduce the MFs and the Fisher matrix we construct to forecast the constraints on MG parameters. In Sec.~\ref{sec:constraints} and Sec.~\ref{sec:halo}, we show our results for the MFs measured from the distributions of dark matter and dark matter halos, respectively. Conclusions and discussions are presented in Sec.~\ref{sec:conclu}.

\section{Models and Simulations}
\label{sec:mod}

An alternative way to solve the enigma of the late-time cosmic acceleration other than introducing dark energy is to modify the left-hand side of the Einstein equation and introduce extra degrees of freedom \cite{2010LRR....13....3D,2012PhR...513....1C,2017PhR...692....1N}. One of the simplest and most popular examples is $f(R)$-gravity, in which the Einstein-Hilbert action is extended to contain a generic function of the Ricci scalar $R$,
\begin{equation} 
S=\int d^4x\sqrt{-g}\left[{R+f(R)\over 16\pi G}\right] +\mathscr{L}_m,
\label{eq:action}
\end{equation}
where $g$, $G$, and $\mathscr{L}_m$ are the determinant of the metric, Newton's constant, and the matter Lagrangian respectively.

There are various functional forms of $f(R)$ proposed in the literature. Here, as a concrete example to study the MG effects on the MFs, we choose the Hu-Sawicki (HS) model \cite{2007PhRvD..76f4004H}, 
\begin{equation} f(R)= -m^2\frac{c_1(R/m^2)^n}{c_2(R/m^2)^n+1}, \end{equation} 
 where $m={H_0}\Omega_M^{1/2}$ is a mass scale, $c_1$ and $c_2$ are free parameters. To match the background expansion of $\Lambda\rm CDM$, they should satisfy $c_1/c_2=6\Omega_{\Lambda}/\Omega_M$, where $\Omega_m$ and $\Omega_{\Lambda}$ are fractional matter and dark energy density parameters respectively. $n$ is the power index and we take $n=1$ in this work. The dynamical degree of freedom in this model is given by the derivative $f_R\equiv {df(R)\over dR}$, which can be taken as a scalar field, often called the scalaron $f_R$. Its present value $f_{R0}$ is conventionally used as the parameter of the model to determine the deviation from GR, with a smaller value of $\left|{f_{R0}}\right|$ representing less deviation from GR.

The enhancement of gravity in the HS model can be treated as a `fifth force' transmitted by the scalaron, which plays a role on all scales. However, the Solar system and laboratory experiments require that GR must be recovered in dense, high-curvature regions. In $f(R)$ gravity, the chameleon mechanism achieves this requirement because the scalaron has a position-dependent mass $m_{\rm eff}$,
\begin{equation}m_{\rm eff}^2\simeq{1\over 3 f_{RR}}\equiv{1\over3}\left|{d^2f(R)\over dR^2}\right|^{-1}, \end{equation}
so that, with a Yukawa-type potential, the fifth force will be exponentially screened when $m_{\rm eff}$ is large. In other words, the effective mass $m_{\rm eff}$ characterizes the Compton wavelength $\lambda_C=m_{\rm eff}^{-1}$, which determines how far the fifth force can propagate. Below the Compton wavelength, gravity will be enhanced with a factor of $4/3$. 

For this work, we use the ``Extended LEnsing PHysics using ANalaytic ray Tracing (\texttt{ELEPHANT})'' dark matter only N-body simulations, which were run using the \texttt{ECOSMOG} codes \cite{li2012ecosmog,li2013exploring}. This code is designed for high accuracy, high resolution, and large volume cosmological simulations for a wide class of modified gravity and dynamical dark energy theories based on the \texttt{RAMSES} code \cite{2002A&A...385..337T}. 
Adaptively refined meshes are created in high-density regions to obtain high resolutions when solving the Poisson and scalaron equations. 
Thanks to the efficient parallelization and the usage of the multigrid relaxation to solve the equations on both the regular domain grid and refinements, high efficiency can be achieved.

The cosmological parameters for GR were chosen 
as the best-fit values from WMAP-9yr 
\cite{2013ApJS..208...19H} results, 
$\{\Omega_b, \Omega_{CDM}, h, n_s, \sigma_8\}$=
$\{0.046$, $0.235$,$0.697$,$0.971$,$0.82 \}$. 
The amplitudes of the primordial power spectrum 
$A_s$ are the same for $f(R)$ and GR simulations, 
hence different gravity models give different $\sigma_8$. 
Each simulation evolves $N_p=1024^3$ dark matter particles 
with mass $m_p=7.8\times 10^{10} h^{-1} \rm M_\odot$ 
in a cubic box with comoving size 
$L_{box}=1024 h^{-1}\rm Mpc$.
We use five independent realizations for 
each gravity model in this work, 
and for both GR and MG models, 
each realization has the same initial conditions (ICs), 
generated by the \texttt{MPGRAFIC} code 
\cite{2008ApJS..178..179P} at initial redshift 
$z_{ini}=49$.

There are three HS models with parameters 
$|f_{R0}|=10^{-4}({\rm F4})$, $10^{-5}({\rm F5})$, 
$10^{-6}({\rm F6})$ simulated in the \texttt{ELEPHANT} 
simulations. Below, we mainly use F5 for our analysis.
To estimate the covariance matrix of the MFs, 
we use $30$ independent $\Lambda\rm CDM$ realizations 
\cite{2021arXiv210902632B} based on \texttt{RAMSES} code 
with similar conditions: $1024^3$ dark matter particles 
in a cubic box with size $L_{box}=1h^{-1}\rm Gpc$, 
and $z_{ini}=49$. These simulations use the same 
level of adaptive mesh refinement as \texttt{ELEPHANT}. 
The cosmological parameters of the simulations are 
$\{\Omega_b,\Omega_{CDM},h, A_s,n_s\} =\{0.048,0.259,
0.68,2.1\times10^{-9},0.96\}$, 
{from which the matter density variance 
parameter can be derived as $\sigma_8=0.824$}.

\section{Methods}
\label{sec:method}
{In this section, we first 
introduce the meaning of the four Minkowksi functionals and
how we measure them. Then we describe the construction of the 
Fisher matrix which we use to obtain the forecast constraints 
on the parameter of $f(R)$ gravity.}

\subsection{Minkowski functionals and their measurement}
\label{subsec:MFs}

According to Hadwiger's theorem \cite{Hadwiger57,rado1959h}, 
the morphological properties of patterns in a $d$-dimension 
space can be fully described by $d+1$ quantities which 
satisfy motion-invariance, additivity, etc, 
namely the Minkowski functionals. 
In this work, the patterns are chosen to be the 
excursion sets characterized by the density 
contrast $\delta$, i.e. regions with density 
contrast $u(\bm{x})~(\equiv{\rho(\bm{x})/\bar{\rho}}-1)$ 
exceeding a specific $\delta$, where $\rho$ and 
$\bar{\rho}$ are density and the mean density of the field, 
respectively. Geometrically, the four MFs $V_i$ with 
$i=0,1,2$ and 3 in 3-dimension space represent the 
excursion sets' volume, their surface area, integrated 
mean curvature, and Euler characteristic $\chi$, 
respectively.

When studying the MFs, one often chooses to use their 
spatial density instead of the MFs themselves to make 
comparisons between samples with different volumes. 
Therefore, we divide the measured MFs by the volume of 
the simulation box. Thus, in this work, $V_0$ is the 
volume fraction occupied by the patterns, 
while $V_1$, $V_2$, and $V_3$ refer to the surface area, 
the integrated mean curvature, and the Euler 
characteristic per unit volume, respectively.

We derive the MG-induced features in the MFs, 
$\Delta{V_i}$, from the differences between the averaged 
values in MG and GR models, which are measured from 5 
independent realizations with $L_{box}=1024h^{-1}\rm Mpc$ 
for each model. We estimate the errorbars using the 
same number of realizations. The procedures to measure 
the MFs are as follows. First, we construct the density 
field from the spatial distribution of the objects 
(dark matter particles or halos) using the cloud-in-cell 
mass-assignment scheme. Next, we smooth the field with a 
Gaussian window function with a smoothing scale $R_G$. 
Then we characterize the excursion sets with the density 
threshold and the MFs are measured as functions of the 
density contrast 
$\delta~(\equiv(\rho-\bar{\rho})/\bar{\rho})$. 
When measuring the MFs, we use the integral method 
developed by \cite{1997ApJ...482L...1S}. We have tested 
the two standard methods they developed, Koenderink 
invariant based on differential geometry and 
Crofton's formula based on integral geometry, 
and find that they give consistent results.

{The MFs are more frequently measured 
as a function of normalized density threshold
$\nu$ or a rescaled volume filling factor $\nu_A$ in the literatures
\cite{1986ApJ...304...15B,1986PThPh..76..952T,2003PASJ...55..911H,
2013MNRAS.435..531C,2013JKAS...46..125P,2019MNRAS.485.4167P,2000astro.ph..6269M}.
The former by definition is $\nu=\delta/\sigma$, where
$\sigma$ is the root mean square of the density fluctuations.
While the latter is defined according to the volume Minkowski
functional $V_0$ and make it match with 
a corresponding Gaussian field: 
$V_0=f_A={1\over{2\pi}}\int_{\nu_A}^{\infty}e^{-t^2/2}dt$,
where $f_A$ is the fractional volume of the field above $\nu_A$.}
{One should not be puzzled by these choices as they 
play different roles. When representing the MFs as a function
of $\nu$, we disregard the variance of the field but
focus more on higher-order properties. If the rescaled
factor $\nu_A$ is chosen, one would further lose 
all Gaussian information due to rescaling, 
meanwhile obtaining other advantages: 
The MFs are invariant under a local monotonic transformation of the
density field and a properly parameterized bias scheme would not introduce
non-Gaussian correction to the MFs \cite{2013MNRAS.435..531C,2000astro.ph..6269M}.} 

\begin{figure*}
   \begin{center}
   \includegraphics[width=1.0\textwidth]{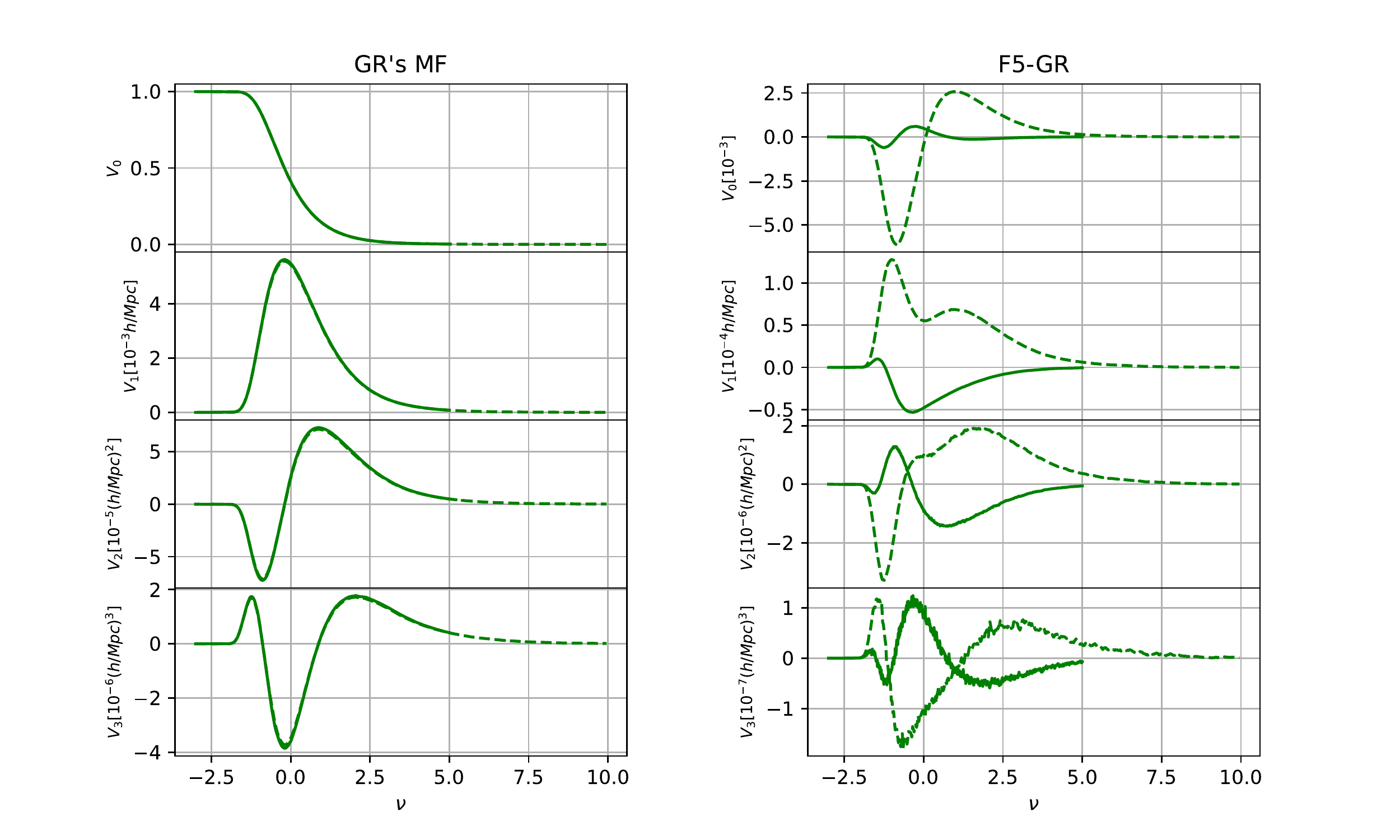}
   \caption{\label{fig:nurho}
{
   The MFs ($left$) and the differences in MFs ($right$) 
   between F5 ($|f_{R0}|=10^{-5}$) and GR 
   as a function of the density contrast 
   $\delta={\rho/\bar{\rho}-1}$ ($dashed$) and the normalized density 
   $\nu={(\rho-\bar{\rho})/\sigma}$ ($solid$) measured 
   with $R_G=10h^{-1}\rm Mpc$ at $z=0$. Where $\rho$, $\bar{\rho}$ and $\sigma$ are the 
   density, mean density and the variance of the density field.
   The dashed lines are rescaled with a factor of $1/\sigma({\rm GR})$ to keep 
   the same $x$-scale with the solid lines.}
   }
   \end{center}
   \end{figure*}

{When the excursion sets are characterized by 
the density contrast $\delta$, all physcical and unphysical 
factors that change the variance of the field would rescale
the $\delta$, hence change the MFs. That is, more informations
will be captured by the MFs as a function of $\delta$ meanwhile
systimatics affect the MFs more.
Normalized density threshold $\nu$ and 
volume filling factor $\nu_A$ are better choices
if one deals with a weakly non-Gaussian case and wants to 
compare their measurements with the theoretical predictions.
However, that is not our point. 
The purpose of this work is using the MFs to distinguish 
the modified gravity and general relativity. 
Thus we follow the works \cite{2020PhRvD.101f3515L, 
WeiLiu2021, liu2023probing,2022PhRvD.105j3028J,2021MNRAS.508.3771L} to directly choose the 
density contrast to define the excursion sets and in which
case we would capture all information including both Gaussian
and non-Gaussian properties of the field.
In figure~\ref{fig:nurho}, we show the measured MG signatures 
when the MFs are chosen as a function of $\delta$ and $\nu$ respectively.
One can find from the
larger amplitudes of $\Delta{V_i}$ for the dashed lines that the difference 
in the variance of the field between 
two gravity models leads to enhancement of the MG signatures.
}

\subsection{Fisher matrix}
\label{subsec:Fisher}
To forecast the constraints on $f(R)$ gravity 
from the Minkowski functionals, we use the Fisher 
matrix technique \cite{1997ApJ...480...22T} to predict error on the modified gravity parameter. Following \cite{2021PhRvD.103l3525R, 2021PhRvD.104j3519L, 2019PhRvD.100l3540W}, we forecast the error on ${\rm log}_{10}(|f_{R0}|)$ to avoid non-zero likelihood for unphysical values of $f_{R0}$. 
Specifically, a Gaussian likelihood is assumed in the ${\rm log}_{10}(|f_{R0}|)$ space around the fiducial value ${\rm log}_{10}(|f_{R0}|)=-5$.
Ignoring the dependence of the covariance matrix on the model parameters, we construct the Fisher matrix as
\begin{equation}
F_{\alpha\beta}={\partial \bm{v}^T\over\partial p_\alpha}\bm{C}^{-1}{\partial \bm{v}\over\partial p_\beta},
\label{eq:fisher}
\end{equation}
where $p_\alpha$ and $p_\beta$ represent model parameters, $\bm{v}$ refers to the vector of observables, which are the MFs for different thresholds and orders and in the most general case measured with different smoothing scales and at different redshifts. $\bm{C}$ is the covariance matrix. Note that, since we have only one parameter here, i.e., $\log_{10}(|f_{R0}|)$, our Fisher matrix is a simple $1\times1$ matrix. Theoretically, the error on $\log_{10}(|f_{R0}|)$ can be estimated by $\sigma_{\log_{10}(|f_{R0}|)} =\sqrt{F^{-1}}$.

To estimate the derivative, we use the following approximation,
\begin{equation}
{\partial \bm{v}\over\partial p}={\bm{\bar{v}}(p)-\bm{\bar{v}}(p^-)\over p-p^-},
\end{equation}
where $\bm{\bar{v}}$ represents the average of the MFs over all realizations of the corresponding model. 
{Given that until very recently there have been very few situations that explore the parameter space of $f_{R0}$ in great detail, most existing simulations use a fixed set of $\log_{10}(|f_{R0}|)$ values, e.g., $\log_{10}(|f_{R0}|)=-5 $, $ -6$. As a result, we follow the common practice as found in but not limited to Ref.\cite{2017MNRAS.466.2402S, 2022MNRAS.514..440R, 2016PASJ...68....4S,2015MNRAS.451.4215Z} and choose $p$ and $p^{-}$ $= -5$ and $-6$ respectively for our derivative estimation. It is essential to note a potential caveat: a 20\% difference in $\log_{10}(|f_{R0}|)$ in our derivative estimation might not meet the desired level of precision for convergence. }
A high-order estimator such as the one used in \cite{liu2023probing} may also provide us with more precise results. However, compared to GR simulations, MG simulations are much more expensive to run, which restricts the possibility of simulating many models with different values of $f_{R0}$ to achieve a more accurate estimation of the derivatives. 
{Due to these reasons, our results can be viewed as intuitive values to illustrate the distinguishability between the two models. For an accurate estimate of the constraint precision, one generally needs to use the Markov Chain Monte Carlo method with likelihoods estimated using full covariance matrices based on proper mock galaxy catalogues, which is also future work~\cite{WeiinPrep} we plan to do.}

Due to the limited number of realizations, we have to use the sub-sample method to get enough samples for the covariance estimation. Our covariance matrix is estimated by
\begin{equation} C_{\mu\nu}={1\over n-1}\sum_{i=1}^n(v^{\rm{obs},i}_\mu-\bar v_\mu)(v^{\rm{obs},i}_\nu-\bar v_\nu), \label{eq:cov}\end{equation}
where $n$ is the total number of sub-boxes, $v^{\rm{obs},i}$ represents the MF measured from the $i-$th sub-box, and the subscript $\mu$ and $\nu$ refer to different thresholds, different orders, different smoothing scales, and different redshifts. For an unbiased estimation, we follow the correction suggested by Hartlap et.al. in \cite{2007A&A...464..399H} and correct the inverse of the covariance matrix with a factor $n-p-2\over n-1$, where $p$ is the degree of freedom of observables. For each order of the MFs, we uniformly choose $21$ density thresholds in the density range $\left[ \delta_{min},\delta_{max} \right]$ for our Fisher analysis. Where $\delta_{min}$($\delta_{max}$) is related to $V_0=0.99(0.01)$. This simple linear spacing bin choice may not be optimal. To further improve the constraints, one can try using non-linear spacing bins or selecting different threshold regions for different orders of MF as done in \cite{liu2023probing}. We have tried selecting the threshold regions according to \cite{liu2023probing}, and found no improvements in our results, likely because our number of bins is limited to be much smaller than that used by \cite{liu2023probing}. Finding the most optimal bin choice is an important topic, and a comprehensive study left as future work~\cite{WeiinPrep}.

For our covariance estimation, we use the 30 \texttt{RAMSES} $\Lambda\rm CDM$ simulations and divide them into 1920 sub-boxes with $L_{box}=250h^{-1}\rm Mpc$ when we study the effects of the smoothing scales, redshifts, and halo bias. Moreover, when studying the effects of cosmic variance, we further divide each sub-box into 8 or 64 smaller sub-boxes with $L_{box}=125 h^{-1}\rm Mpc$ or $62.5 h^{-1}\rm Mpc$. Then we estimate the covariance matrices from these subsamples for the three subbox volumes. We test the convergence of our covariance matrix in Appendix~\ref{append:covariance}.

{ Since we choose F5 as the fiducial model and estimate the covariance matrix with \texttt{RAMSES} runs which have both different cosmological parameters and gravity, the concern is also raised about how the covariance matrices are dependent on these factors. To test the issue, we take a comparison between the results obtained from \texttt{RAMSES} subsamples and those from \texttt{ELEPHANT} subsamples. The results are shown in Appendix ~\ref{covariance_parameters}  }
%

\section{Constraints from the MFs of Dark Matter}
\label{sec:constraints}
In this section, we quantify the MG information content encoded in the MFs of the dark matter density field by the Fisher matrix technique mentioned above. It is valuable to evaluate how survey volume affects the constraint on the MG parameters. We first discuss how cosmic variance will affect our constraint in Sec.~\ref{sec:volume}. Next, considering the scale-dependence of the MG effects, we study the MG signal encoded in the MFs measured with different smoothing scales $R_G$ in Sec.~\ref{sec:ss}. Then in Sec.~\ref{sec:redshift}, we investigate the MG effects at different redshifts. Finally, in Sec.~\ref{sec:combine}, we try to combine the observables measured with different smoothing scales and at different redshifts to tighten the constraints.

\subsection{Cosmic variance}
\label{sec:volume}

\begin{figure}
\centering
\includegraphics[width=1.0\linewidth]{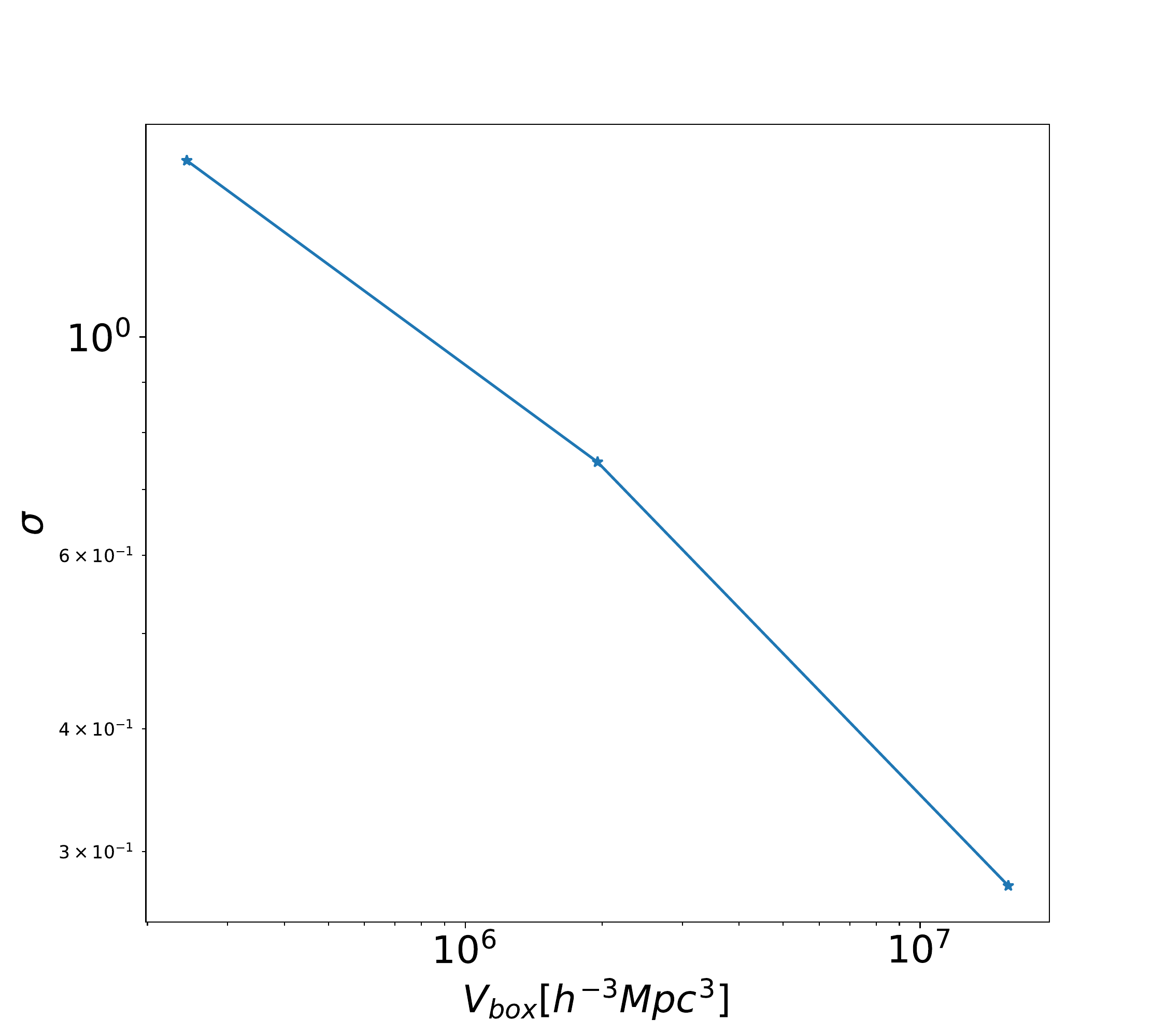}
\caption{\label{fig:f5ssize} Constraints on ${\rm log}_{10}(f_{R0})$ from the MFs estimated by the Fisher matrix technique, as a function of the forecasted box volume $V_{box}$. The MFs are measured from the dark matter density field smoothed with a Gaussian window of size $R_G=5h^{-1}\rm Mpc$ at $z=0$. The three data points from left to right are obtained with the covariance matrices estimated from subsamples with $V_{box}=62.5^3h^{-3}\rm Mpc^3$, $125^3h^{-3}\rm Mpc^3$ and $250^3h^{-3}\rm Mpc^3$ respectively. }
\end{figure}

The cosmic variance is related to the finite volume of a galaxy survey. It is important for both cosmological parameter estimation and future survey design, and it has been extensively studied for various LSS statistics \citep{1994A&A...281..301C, 1996ApJ...470..131S, 2011ApJ...731..113M, 2003ApJ...584..702H, 2012MNRAS.423..909C, 2010MNRAS.407.2131D}. Because our main results in this work are based on analysis of sub-boxes with $L_{box}=250h^{-1}\rm Mpc$, while the size of redshift surveys nowadays is often much larger than this value, it is necessary to quantify how the cosmic variance will affect the constraints from the MFs. 

To do this we separately estimate the covariance matrix from the subsample sets with different box volumes. These subsamples are optimistically assumed to only receive the influence of the box volume, and the MFs are measured from each box with $R_G=5h^{-1}\rm Mpc$ at $z=0$. 
We show in Fig.~\ref{fig:f5ssize} the constraints as a function of box volume $V_{box}$. {The three data points from left to right represent constraints from sub-boxes with $V_{box}=62.5^3$,$125^3$, and $250^3~h^{-3}\rm Mpc^3$ respectively. We find the value of $\sigma_{{\rm log}_{10}(|f_{R0}|)}$ decreases by a factor of 2.0 when the box volume increases 8 times from $62.5^3h^{-3}\rm Mpc^3$ to $125^3h^{-3}\rm Mpc^3$, and by another factor of 2.7 when the volume increases from $125^3h^{-3}\rm Mpc^3$ to $250^3h^{-3}\rm Mpc^3$. } 

To understand this scaling, one can assume the boxes in a sub-sample set are statistically independent and identically distributed. When we divide $N$ large boxes into $nN$ sub-boxes, the statistics $S$ with additivity satisfy $S_{large}=nS_{sub}/n=S_{sub}$, and their standard deviations satisfy $\sigma_{large}=\sqrt{n}\sigma_{sub}/n=\sigma_{sub}/\sqrt{n}$. Hence one can expect $\sigma_{f_{R0}}$ should improve by a factor of $\sqrt{8}$ if the volume is enlarged by 8 times. However, the basic assumption in the subsample method, that all subsamples are fully independent of each other, is not really satisfied, which leads to systematical bias when estimating the covariance \cite{lacasa2017inadequacy}. And the bias might be more significant for subsamples with smaller volumes because of the stronger correlations between small-scale structures. This may be the reason why the constraint is not strictly scaled as $1/\sqrt{V}$. There are methods to study the effects of the sample variance more accurately, such as by running more independent simulations with different volumes, or by Monte Carlo sampling for a large survey. We will consider them in future work.

This scaling with $V$ can in principle be extended to other cosmological parameters, indicating that with the huge volume of future surveys, we can obtain much more accurate constraints on both MG and other cosmological parameters from the MFs. For example, the DESI spectroscopic survey \cite{2016arXiv161100036D} covers a sky area of $\sim14000\rm deg^2$, and is expected to detect objects in the redshift range $z\lsim1.9$, corresponding to a comoving volume $\sim 60 h^{-3}\rm Gpc^3$. Both of them are much larger than the volume $V_{box}=1.56\times10^{-2}h^{-3}\rm Gpc^3$ that we use to obtain our main predictions. 


\subsection{Smoothing scales}
\label{sec:ss}

\begin{figure*}
\begin{center}
\includegraphics[width=1.0\textwidth]{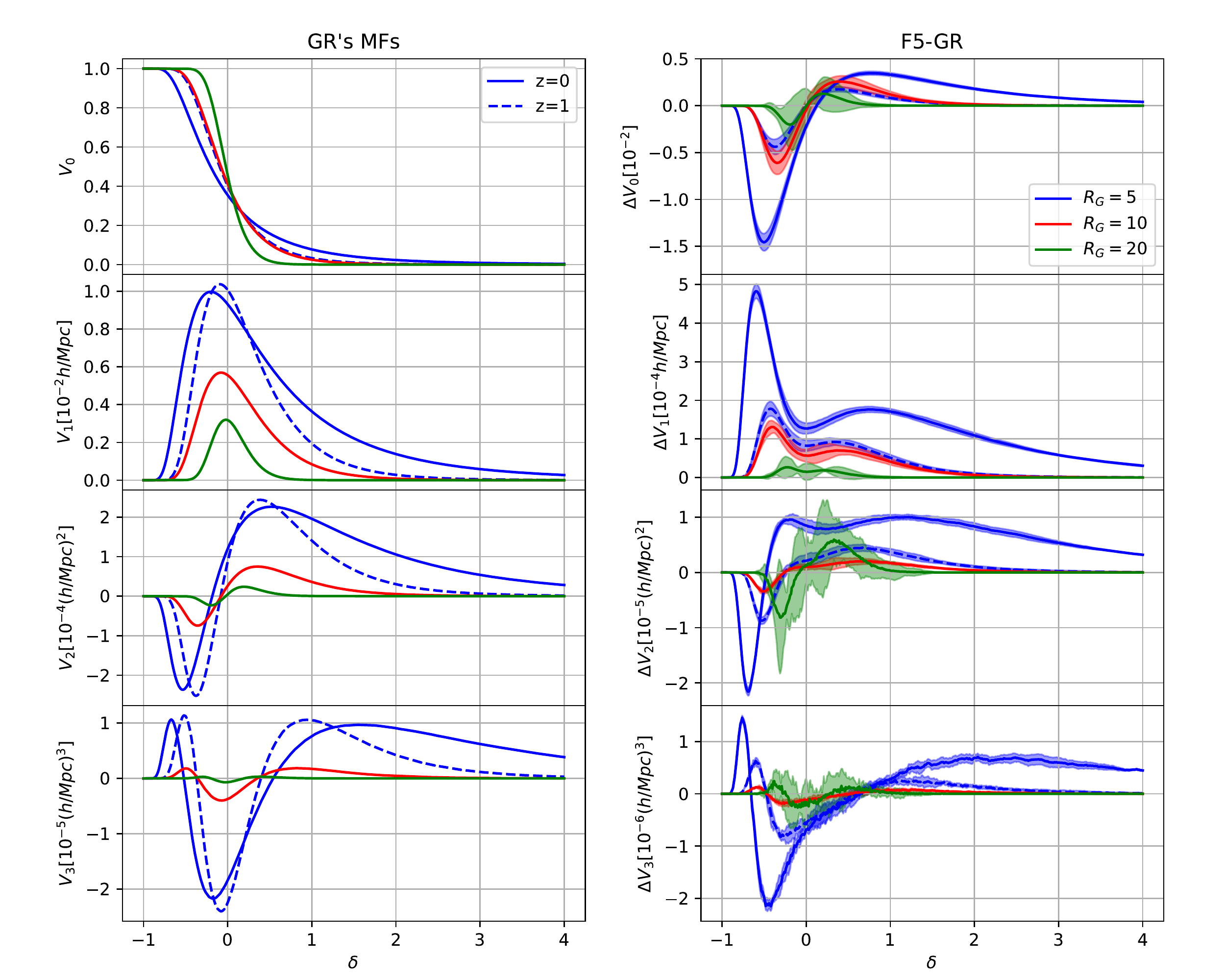}
\caption{\label{fig:mfs} 
The MFs ($left$) and the differences in MFs ($right$) 
between F5 ($|f_{R0}|=10^{-5}$) and GR 
as a function of the density contrast 
$\delta={\rho/\bar{\rho}-1}$. 
The MFs are measured from the dark matter density field 
at $z=0$ ($solid$) and $z=1$ ($dashed$). 
We smooth the field by a Gaussian window function with 
scale $R_G= 5,~ 10,~20$ (unit: $h^{-1}\rm Mpc$)
({for $z=1$ we only show the results with
$R_G=5h^{-1}\rm Mpc$
because the results at the two redshifts changes similarly with 
changing $R_G$}), respectively. 
{The shaded regions represent the error bands.} 
These errors are estimated from five independent 
realizations with volume $V_{box}\simeq1.07h^{-3}\rm Gpc^3$. 
$\Delta{V_2}$, $\Delta{V_3}$ and their associated errors 
are artificially enlarged by a factor of $20$ 
when $R_G=20h^{-1}\rm Mpc$ in the figure 
for ease of visualization. }
\end{center}
\end{figure*}

Before the measurement of the MFs, 
the density field is often smoothed by a Gaussian window function 
with a smoothing scale $R_G$ to reduce the shot noise 
(e.g., \cite{1997ApJ...482L...1S, 2013MNRAS.429.2104D, 
2017PhRvL.118r1301F, 2020PhRvD.101f3515L}). 
For example, in Ref \cite{2020PhRvD.101f3515L},
it was demonstrated that choosing an optimal 
value of $R_G$ can improve the signal-to-noise (S/N) 
ratio of neutrino signatures when employing MFs to 
study the impact of massive neutrinos on LSS. 
Furthermore, Ref \cite{WeiLiu2021} quantified 
the cosmological constraint obtained 
from the MFs of the simulated dark matter (and neutrino) 
field with $R_G=5h^{-1} \rm Mpc$ and $10h^{-1}\rm Mpc$, 
and found their combination can improve the results. 
Similar to neutrinos, MG's effects on LSS are also 
scale-dependent. Hence in this work, we also try 
different $R_G$s to study the scale-dependent 
effects on the MFs of LSS from MG.

We show in Fig.~\ref{fig:mfs} the 
MF $V_i$ measured with $R_G=5h^{-1} \rm Mpc$, 
$10h^{-1}\rm Mpc$ and $20h^{-1}\rm Mpc$ at $z=0$ 
from the GR and F5 models and the difference 
$\Delta{V_i}$ between the MFs of the two models. 
The values of $V_i$ are the means measured from 
the five independent simulations with 
$L_{box}=1024 h^{-1}\rm Mpc$, 
while the error bars represent their standard deviations. 
These are shown as a function of dark matter 
density contrast $\delta$ for 
{$\delta\in[-1,4]$}. 
The three $R_G$ choices from the smallest to 
the largest values roughly represent a non-linear, 
quasi-linear, and linear scale respectively. 
We will briefly interpret the curves of $V_i$ 
measured with {$R_G=15h^{-1}\rm Mpc$}
from the halo field 
and compare them to results for the matter density field 
in Sec.~\ref{sec:halo}, as the results with 
$R_G=5h^{-1}\rm Mpc$ have been discussed in detail 
in \cite{2017PhRvL.118r1301F}. In this section, 
we only focus on the difference between 
the curves with different $R_G$.

We find from the curves of $V_0$ that when $\delta\gsim0$ 
the excursion sets occupy a smaller fractional volume for 
larger $R_G$s. And the fractional volume occupied by the 
excursion sets is larger when $\delta\lsim0$, 
which indicates a smaller fractional volume occupied 
by their complements, i.e., the under-dense regions. 
These results are consistent with an overall 
smaller density fluctuation for a larger $R_G$, 
for the process of smoothing erases structures 
with scales smaller than $R_G$. 
Other orders of $V_i$ have similar trends with 
lower amplitudes when we increase $R_G$ due to 
the same reason. It's worth mentioning that as $R_G$ 
increases, all $V_i$ curves tend to approach the 
Gaussian case.

In the right panel,  we find the curves of 
$\Delta{V_i}$ also have similar shapes for 
different $R_G$s but lower amplitudes for larger $R_G$s. 
The signal-to-noise ratio also decreases 
when we increase the smoothing scale. 
The Compton wavelength for F5 is 
$\sim 7.5h^{-1}\rm Mpc$ at $z=0$. 
One can expect more MG signals below the scale. 
Note that the process of Gaussian smoothing 
does not erase all information on scales smaller than $R_G$,
but gives an exponentially suppressed weight to 
structures on these scales. 
This is the reason why we can still find MG signals for 
$R_G$ larger than the Compton wavelength.

To quantify the influence of smoothing scales, 
we estimate the parameter error 
$\sigma_{{\rm log}_{10}(f_{R0})}$ by the Fisher matrix, 
which is constructed from the MFs of the dark matter field 
with four $R_G$ choices $R_G= 5,~10,~15,$ and 
$20~h^{-1}\rm Mpc$ at $z=0$. 
The covariance matrix is estimated from the 1920 subsamples 
with size $L_{box}=250h^{-1}\rm Mpc$. 
We show the results in Table.~\ref{tab:constraint_value}. 
The constraint becomes monotonically worse as we 
increase $R_G$,  consistent with the findings in Fig.~\ref{fig:mfs}.
{The best constraint $\sigma_{{\rm log}_{10}(|f_{R0}|)}=0.28$ 
is obtained with $R_G=5h^{-1}\rm Mpc$.} 

\begin{figure}[htbp] 
   \centering
   \includegraphics[width=1.0\linewidth]{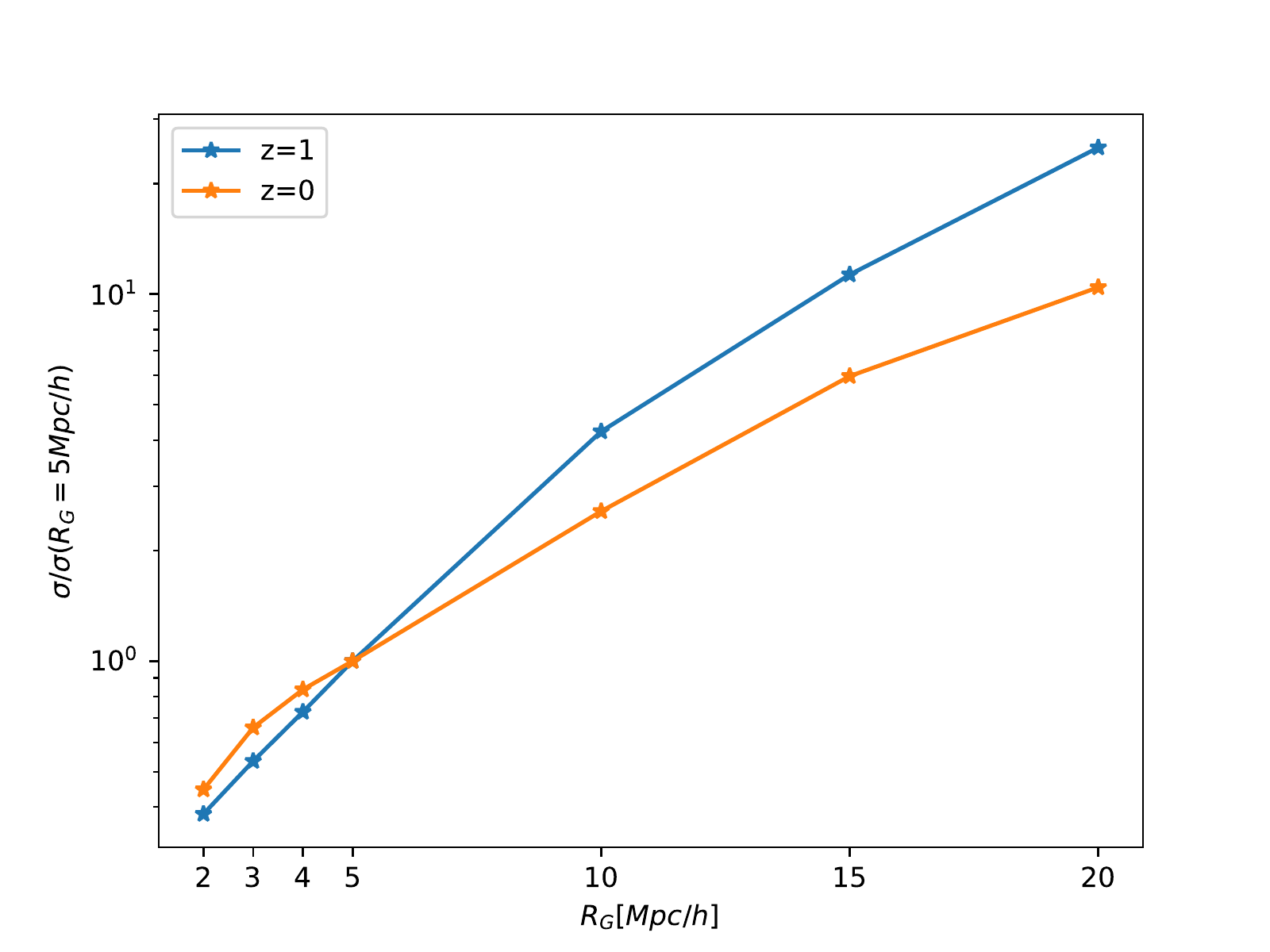} 
   \caption{ {Ratio of the constraint on ${\rm log}_{10}(f_{R0})$, 
   $\sigma_{{\rm log}_{10}(f_{R0})}$,  
   obtained from the MFs measured with Gaussian smoothing 
   scale $R_G$ to that with $R_G=5h^{-1}{\rm Mpc}$. 
   The blue and orange lines represent the ratios at $z=1$ 
   and $z=0$ respectively.}}
   \label{fig:smallscale}
\end{figure}

We propose $R_G=5h^{-1}\rm Mpc$ as a realistic choice 
that is achievable for next-generation galaxy surveys 
\cite{2016arXiv161100036D}. 
However, as a theoretical analysis, 
we can push $R_G$ onto smaller scales to check 
if there is an optimal scale where we could obtain 
the best constraint. 
In Fig.~\ref{fig:smallscale} we show the ratio of the 
constraint obtained from the MFs measured with Gaussian 
smoothing scale $R_G$ to that with $R_G=5h^{-1}{\rm Mpc}$. 
Note that due to the high resolution of the N-body simulation, 
even if we smooth the field with $R_G$ as small as 
$R_G=2h^{-1}{\rm Mpc}$, the root mean square of the 
smoothed density field is still much larger than shot noise. 
The result shows a monotonical trend as the constraint 
becomes better when $R_G$ is decreased. 

\subsection{Redshifts}
\label{sec:redshift}


The motivation to investigate the MFs of LSS 
at different redshifts is that the effects of 
the modification to gravity are generally redshift dependent 
and accumulate over time. In $f(R)$ the Compton wavelength 
$\lambda_C$ that determines how far the scalaron 
can propagate is a function of $z$. 
The large-scale distribution of tracers at different 
redshifts can provide us with complementary information. 

We show the MFs $V_i$ measured from the dark matter density 
field with $R_G=5h^{-1}\rm Mpc$ at $z=1$ 
and the corresponding differences $\Delta{V_i}$ 
in the MFs between two models {as dashed lines in 
Figure.\ref{fig:mfs}}.

The morphological properties for a Gaussian random field 
are statistically equivalent for overdense and 
under-dense regions. Thus in the Gaussian case, 
the curves of $V_0$ and $V_2$ are central symmetric about 
the origin, while the curves of $V_1$ and $V_3$ are 
symmetric about $\delta=0$. Because of non-linear evolution 
on small scales, these curves deviate from the Gaussian case.
And there are greater deviations at $z=0$ compared to $z=1$, 
due to more non-linear evolution at low redshift. 
From the curve of $V_0$, we find that the excursion sets 
occupy a smaller fractional volume at $z=1$ for 
$\delta\gsim0$. Also, they occupy a larger fraction of 
total volume when $\delta\lsim0$, hence their complements, 
the under-dense regions, occupy a smaller volume fraction. 
These findings reflect a relatively smaller density 
fluctuation at $z=1$. To confirm the result, 
we measure the root mean square of the smoothed density 
field by $r.m.s.=\sqrt{<(\rho-\bar\rho)^2>}$, 
where $\rho$ and $\bar\rho$ are the density and 
mean density of the field respectively, 
and the bracket represents averaging over total grids.  
We find the $r.m.s.$ is $\sim40\%$ smaller 
at $z=1$ than that at $z=0$, 
which is consistent with our findings. 

From the right panels of Fig.~\ref{fig:mfs}, 
we find that all curves of $\Delta{V_i}$ have similar shapes 
for the two redshifts, but the amplitudes are lower at 
the higher redshift. Considering the curves of the MFs 
in the left panels, one can easily find the relative 
difference of $\Delta V_i/V_i$ 
between the two models is more significant at low redshift. 
That is expected because the Compton wavelength 
at $z=1$ is $\sim3.6h^{-1}\rm Mpc$, 
which is smaller than its value at $z=0$, 
and the modification to gravity is stronger at low redshift.

We also perform the Fisher matrix analysis to 
quantify the constraining power of the MFs at $z=1$, 
and the results are presented in 
Table.~\ref{tab:constraint_value}. 
We find that, compared to the results at z=0, 
the constraints are worse at higher redshift for each $R_G$. 
This is consistent with the findings in Fig.~\ref{fig:mfs}. 
Additionally, it keeps the trend that the constraint is 
better for smaller $R_G$, and the best constraint 
is {$\sigma_{{\rm log}_{10}(|f_{R0}|)}=0.57$} 
with $R_G=5h^{-1}\rm Mpc$. We can also find from 
Fig.~\ref{fig:smallscale} that the trend when tracing to 
smaller scales is the same as the result at $z=0$, 
while the curve at $z=1$ is a little steeper.

\subsection{Combined constraints}
\label{sec:combine}

\begin{table}[h]
   \centering
   
      \begin{tabular}{rcllll}  
      
         \toprule\hline
      ~   &$R_G[h^{-1}\rm Mpc]$ &5  & 10 & 15 &20   \\
         \hline
         Matter, $\mathcal{R}$& $z=0$  & 0.28   & 0.73   & 1.69   &  2.96  \\
         ~ &$z=1$     & 0.57   & 2.41   & 6.44   &  14.20 \\
         ~& $C_0$ &\multicolumn{4}{c}{0.26}\\
            ~ & $C_1$ &\multicolumn{4}{c}{0.47}\\
            ~  &$C_0+C_1$& \multicolumn{4}{c}{0.23}\\
         
         \hline
         \hline
         Halo, $\mathcal{R}$&\multicolumn{1}{c}{$z=0,\mathcal{H}_1$} &--&--&1.10&1.39\\
         ~&\multicolumn{1}{c}{~~~~~~~~$\mathcal{H}_2$} &--&--&1.12&1.41\\
         ~&\multicolumn{1}{c}{~~~~~~~~$\mathcal{H}_{2m}$}&--&--&1.11&1.45\\
         ~&\multicolumn{1}{c}{~~~~~~~~$\mathcal{H}_3$} &--&--&1.25&1.39\\
         \hline
         Halo, $\mathcal{Z}$&\multicolumn{1}{c}{$z=0,\mathcal{H}_1$} &--&--&1.24&1.67\\
         ~&\multicolumn{1}{c}{~~~~~~~~$\mathcal{H}_2$} &--&--&1.24&1.53\\
         ~&\multicolumn{1}{c}{~~~~~~~~$\mathcal{H}_{3}$}&--&--&1.36&1.73\\
         \hline
         Halo, $\mathcal{R}$&\multicolumn{1}{c}{$z=1,\mathcal{H}_1$} &--&--&1.09&1.33\\
         ~&\multicolumn{1}{c}{~~~~~~~~$\mathcal{H}_2$} &--&--&1.23&1.56\\
         ~&\multicolumn{1}{c}{~~~~~~~~$\mathcal{H}_{3}$}&--&--&0.95&1.09\\
         \hline
         Halo, $\mathcal{Z}$&\multicolumn{1}{c}{$z=1,\mathcal{H}_1$} &--&--&1.35&1.45\\
         ~&\multicolumn{1}{c}{~~~~~~~~$\mathcal{H}_2$} &--&--&1.49&1.70\\
         ~&\multicolumn{1}{c}{~~~~~~~~$\mathcal{H}_{3}$}&--&--&1.01&1.29\\
         \hline

         \bottomrule
      
       \end{tabular}

   \caption{{Value of $\sigma_{{\rm log}_{10}(|f_{R0}|)}$ forecasted 
   using the Fisher matrix of MFs based on N-body simulations. 
   The results are obtained from the dark matter field and 
   from three halo populations with their properties listed 
   in table~\ref{halo_population_properties}
   at two redshifts $z=0$ and $1$ (for $z=0$, additional halo
   population $\mathcal{H}_{2m}$ which keeps same lower mass limit for GR and $f(R)$
   is also considered) and in real ($\mathcal{R}$)
   and redshift ($\mathcal{Z}$) space. 
   $C_0$ and $C_1$ represent the results obtained by combining 
   the MFs measured with $R_G=5$, 10, and 20$h^{-1}\rm Mpc$ 
   at $z=0$ and $1$, respectively. While $C_0+C_1$ represents
   the result with combining $C_0$ and $C_1$.
   For each order of MFs, 21 threshold bins
   are chosen. We estimate the covariance matrix from the subsample
   of \texttt{RAMSES} realizations with their volume 
   $V_{s}=1.56\times10^{-2} h^{-3}{\rm Mpc}^3$.}
    }
   \label{tab:constraint_value}
   
\end{table}

In previous subsections, we have evaluated the constraining power of the MFs measured with specific $R_G$ and at specific redshift from the simulated dark matter field with a typical subsample volume $V_{box}=1.56\times10^{-2} h^{-3}\rm Gpc^3$. The volume roughly corresponds to a redshift survey with $z_{max}=0.1$ and sky-coverage $\sim6000\rm deg^2$. We note that the next-generation redshift surveys can be much larger than this choice both for the redshift and sky coverage, hence we can expect more precise results. 

In addition, we can consider the combination of the observables measured with different smoothing scales and at different redshifts to improve our constraints. Since the smoothing process suppresses the structures with scales smaller than $R_G$, the combination of observables with different $R_G$ is equivalent to collecting multi-scale information. We assume no correlation between the observables measured at different redshifts because of the large separation between the two redshifts we consider.

$C_0$, $C_1$ and $C_0+C_1$ in Table.~\ref{tab:constraint_value} show the constraints obtained with the combined observables.
We find that the improvement is not much significant in both two redshift cases when combining three smoothing scales and two redshifts.   
There are two reasons why the improvements after combining different smoothing scales are not significant: first, the correlations between the MFs with different $R_G$s are strong, and then the MG effects are not significant on the large scales, i.e., $R_G$ is $10$ or $20h^{-1}\rm Mpc$, but on small scales as one can find in Fig.~\ref{fig:mfs}. These findings indicate we should try our best to come to small scale and low redshift in future surveys if we want to obtain a good result.

\section{MFs of Dark Matter halos}
\label{sec:halo}

In this section, 
we study the MG signals encoded in the MFs of 
halo distribution and compare them with 
the results for the dark matter. 
We briefly describe how we construct the halo catalogue 
in Sec.~\ref{sec:halocat}. 
{Due to the low number density of the halo catalogue,
we also discuss how the shot noise affects our results in 
Sec.~\ref{subsec:shot noise}}
Then in Sec.~\ref{sec:halo mf}, 
we present our analysis of how the halo bias 
affects the MG signals contained in the MFs. 
Finally, we quantify the constraining power on the MG parameter 
from the MFs of the halo distribution in Sec.~\ref{sec:halo con}. 

\subsection{Construction of the halo catalogue}
\label{sec:halocat}

\begin{figure}
   \centering
   \includegraphics[width=1.0\linewidth]{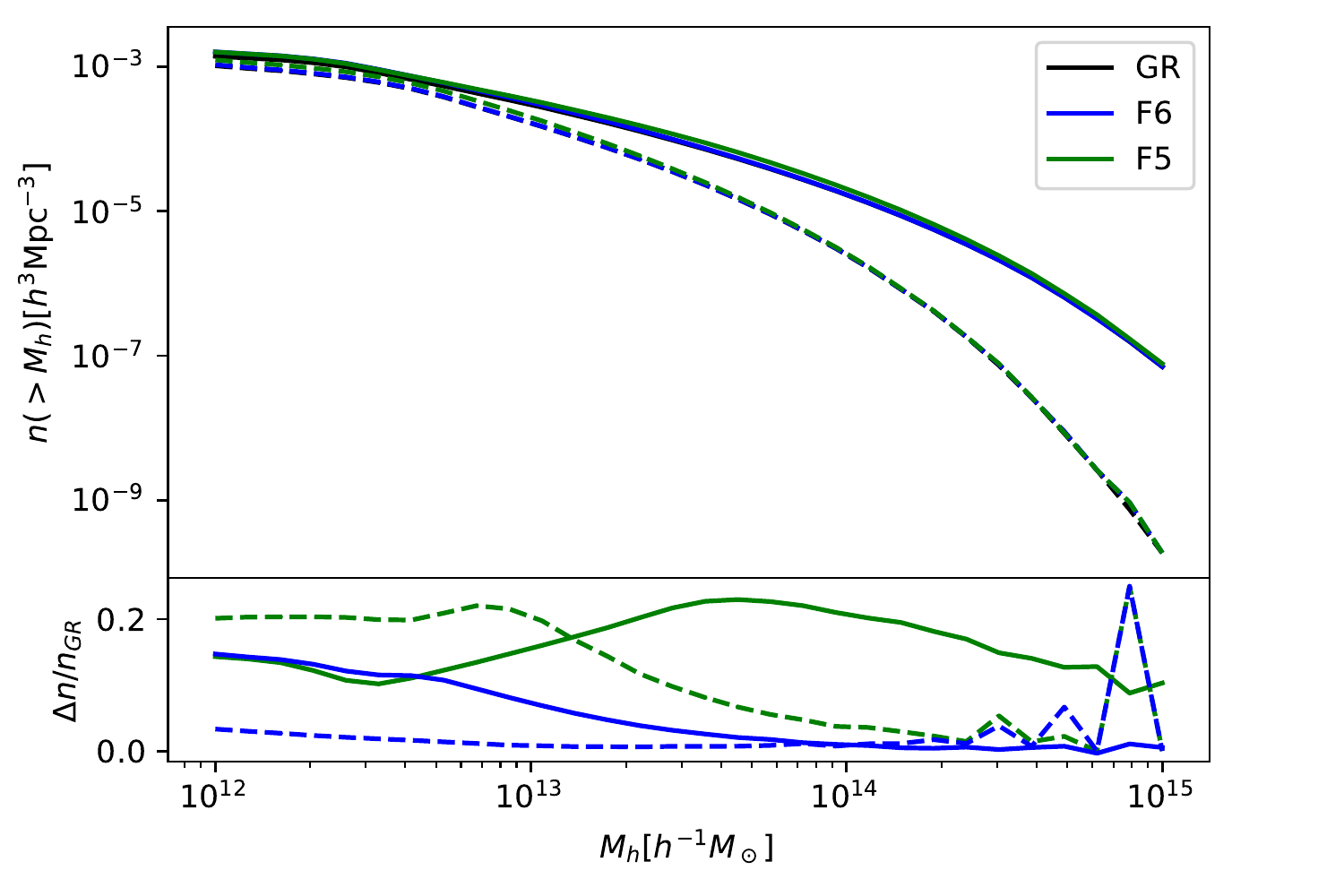}
   \caption{$upper$: Halo mass functions of three Models
    at $z=0$ ($solid$) and $z=1$ ($dashed$). $lower$: 
    Relative difference in the halo mass function between
    modified gravity and GR.}
   \label{halo_mass_function}
 \end{figure}

{Dark matter halos serve as
 the foundational units of large-scale structures 
 and they are the hosts of galaxies. 
 Hence studying their statistical properties, 
 such as clustering and abundance, 
 is important in understanding the nature of gravity. 
 The halo catalogues are constructed using the 
 public \texttt{ROCKSTAR} \cite{2013ApJ...762..109B} algorithm,
which is a friend-of-friend halo finder in 6D phase space.
We define the mass of a halo as $m_{200c}$, 
the mass within a sphere of radius $r_{200c}$, 
which is the radius in which the mean overdensity is 
200 times the critical density of the universe. }

\begin{table}[htbp]
   \centering
   \caption{Properties of the halo populations used in this
   work. We list the minmum halo mass $M_{min}$ [$10^{12}h^{-1}M_\odot$] used to 
   obtain a required number density $\bar{n}$ [$10^{-3}h^3{\rm Mpc}^{-3}$] for a given redshift 
   and gravity model. We also measure a scale-independent bias 
   for these halo populations and show them in the table as a reference 
   to help understand our MFs results.}
   \label{halo_population_properties}
   \begin{tabular}{ccc|ccc|ccc}
      \toprule
      \hline
      ~&~&~&\multicolumn{3}{c}{$M_{min}$}&\multicolumn{3}{c}{bias}\\
      \hline
     ~& ~ &  $\bar{n}$ & GR & F6 & F5& ${\rm GR}$ & F6 &F5 \\
      \hline
      $z=0$&$\mathcal{H}_1$  &  $1.0$ &2.57&3.04&2.96&1.26&1.23&1.23\\
      ~    &$\mathcal{H}_2$  &  $0.7$ &3.98&4.45&4.45&1.36&1.33&1.34\\
      ~    &$\mathcal{H}_3$  &  $0.4$ &7.33&7.96&8.50&1.51&1.47&1.48\\
      \hline
      $z=1$&$\mathcal{H}_1$  &  $1.0$ &1.09&1.17&1.87&2.12&2.10&2.03\\
      ~    &$\mathcal{H}_2$  &  $0.7$ &2.65&2.81&3.43&2.21&2.20&2.15\\
      ~    &$\mathcal{H}_3$  &  $0.4$ &5.07&5.15&5.85&2.47&2.47&2.37\\
      \bottomrule
   \end{tabular}
\end{table}

{Fig.~\ref{halo_mass_function} 
shows the halo mass function, 
which is the dark matter halo number density 
as a function of their mass, 
for two modified gravity models and GR at $z=0$ and $1$. 
Due to the enhancement of the gravity in $f(R)$ scenarios, 
both of the two $f(R)$ models predict more halos 
at almost all masses compared with the GR case 
\cite{2015PhRvL.115g1306H,
2012MNRAS.421.3481L,
2017MNRAS.467.1569A,
2018MNRAS.479.4824H} except the high mass end for F6.
This is because the chameleon mechanism works efficiently
for these halos and the enhancement of the gravity is 
suppressed. }


{To understand the MG signatures 
contained in the MFs of the halos, 
we set the same lower halo mass cut, 
$M_{min}=3.98\times10^{12}h^{-1}M_{\odot}$, for $f(R)$ 
and GR scenarios (hereafter, $\mathcal{H}_{2m}$). 
However, considering a realistic case 
where the number density of observables 
in a galaxy survey is always fixed, 
we also set a fixed number density (hereafter, $\mathcal{H}_2$), 
$\bar{n}=7\times10^{-4}h^{3}{\rm Mpc}^{-3}$, 
which is the number density of the GR model in $\mathcal{H}_{2m}$,
for each of the halo samples. 
According to Figure~\ref{halo_mass_function} 
the halo mass cut should be higher 
in the $f(R)$ case. 
We also consider the halo populations with a fixed but 
higher/lower halo number density than $\mathcal{H}_2$
to study how different halo populations affect the MG signals
we detect.
The number density of the two populations are 
$\bar{n}=1.0\times10^{-3}h^{3}{\rm Mpc}^{-3}$ (hereafter, $\mathcal{H}_1$) 
and $4.0\times 10^{-4}h^{3}{\rm Mpc}^{-3}$ (hereafter, $\mathcal{H}_3$), respectively.
Because we only need the choice I to understand the full
influence of the MG effects, we measure the MFs from it at $z=0$.
But for the other three halo populations, we consider
both two redshifts $z=0$ and $1$ to study how the MG 
effects evolve with time.
}.

{
The lower number densities have been chosen to be 
representative of the samples observed in the last-Stage
galaxy surveys such as SDSS \cite{sdss2000} and BOSS \cite{2017MNRAS.470.2617A}, and the high
number density is expected in the current, e.g., DESI survey \cite{2016arXiv161100036D}.
The minimum halo mass for each halo sample at a given redshift
is listed in table~\ref{halo_population_properties}
}

{
We construct the halo number density field using the cloud-in-cell (CIC)
mass assignment scheme with the grid unit $4h^{-1}\rm Mpc$. 
We have tested mass assignment schemes with different 
accuracy levels, NGP, CIC, TSC and PCS using 
the public \texttt{Pylians} code and find that they provide us with
consistent results if $R_G$ is larger than the grid unit.
}

\subsection{Choice of smoothing scale}
\label{subsec:shot noise}
{The estimation accuracy of the MFs is expected to increase in a 
fixed volume with we choosing a smaller $R_G$ because of the 
increasing number of structures with a typical scale $R_G$. However,
$R_G$ should be limited by a minimum value to satisfy some criteria
\cite{2002ApJ...580..663H,2003PASJ...55..911H}. First, $R_G$ should be 
larger than twice the grid cell, which is related to Nyquist resolution frequency
and can also avoid the systematics caused by the mass assignment scheme. 
Second, $R_G$ should be larger than the mean separation of the objects, to 
avoid treating a single object as a structure. That limits our minmum value of 
$R_G$ as $15h^{-1}\rm Mpc$.
}

{
Third, the shot noise should be properly handled after the density field is
smoothed with $R_G$. We follow the discussions in \cite{liu2023probing} to 
show the validity of our $R_G$ choices. Under the Gaussian limit, 
the four MFs of a biased (halo/galaxy) field with a linear bias $b$ 
should be determined by,
\begin{equation}
   \label{biased sigma}
   \left({\hat{\sigma}_0\over\hat\sigma_1}\right)^2
   = {b^2\sigma_0^2+((4\pi)^{3/2}R_G^3\bar{n})^{-1} \over
   b^2\sigma_1^2+
   {3((16\pi)^{3/2}R_G^5 \bar{n}})^{-1}},
\end{equation}
where $\sigma_i$ is the $i$-th moment of the field, which with 
and without hat represent the biased and matter field respectively. $\bar{n}$
is the mean number density of the biased tracers.
}

\begin{figure}
   \centering
   \includegraphics[width=1.0\linewidth]{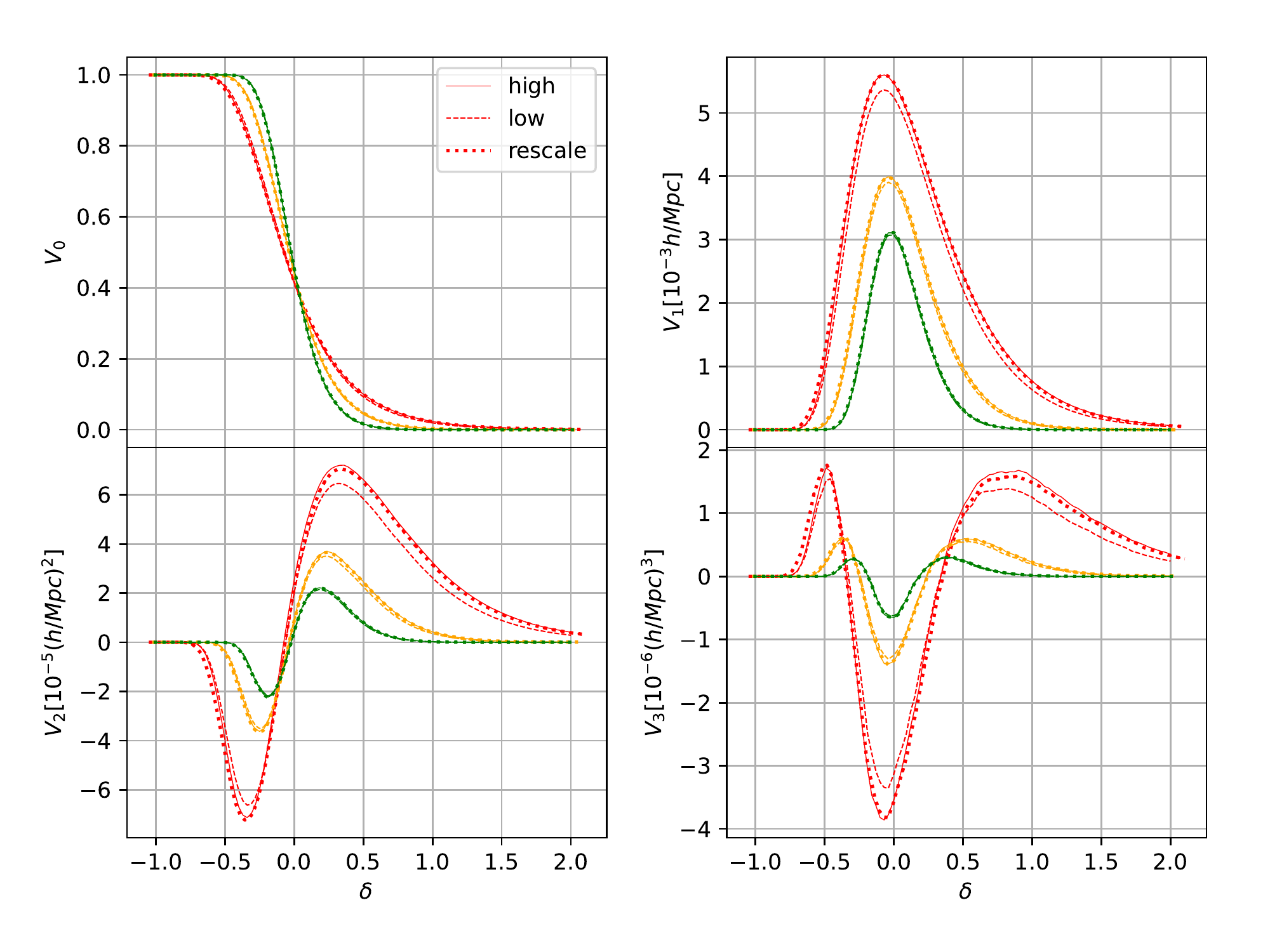}
   \caption{\label{fig:shot_noise}  {The MFs measured
   from the COLA realizations with the number density of dark matter
   particles $n=1.7\times10^{-2}$ ($high$, $solid$ lines) and 
   $9\times 10^{-4} h^{3}{\rm Mpc}^{-3}$ ($low$, $dashed$ lines).
   The colors $red$, $orange$ and $green$ represent the MFs 
   measured with $R_G=10$, $15$ and $20h^{-1}{\rm Mpc}$. The results
   after correcting the shot noise effects are shown as the $dotted$ lines.}}
\end{figure}

{
Therefore, the effect of shot noise for
a biased field with number density $\bar{n}_b$ can be 
equivalent to that for a matter field with number density
$\bar{n}_m\sim \bar{n}_b\times b^2$. We approximately 
measure the bias for our three halo populations by,
\begin{equation}
   b^2=\left< P_h(k)/ P_m(k) \right>,
\end{equation}
where the bracket represents averaging over all $k$ with $k<0.5$.
The values of halo bias are also 
listed in table~\ref{halo_population_properties}.
The values of the galaxy bias can be 
found in table 1 in \cite{2019MNRAS.485.2194H}.
We thus choose the smallest $\bar n_m\sim9\times 10^{-4}$
which is related to $\mathcal{H}_3$ at $z=0$, as a representative to study 
how the shot noise affects our results.
To do this, we use the public \texttt{MG-PICOLA} code \cite{2017JCAP...08..006W} to fast run realizations
with the same cosmology and simulation parameters but two different
resolutions with the number density of the dark matter particles 
$n=1.7\times10^{-2}$ ($high$) and $n=9\times10^{-4}$ ($low$) 
$h^{3}{\rm {Mpc}^{-3}}$.
The results are shown in figure~\ref{fig:shot_noise} as 
solid and dashed lines for high and low resolutions respectively.
}

{
One can find that all $V_i$s show 
good consistency between the high and low-resolution results 
for the two larger $R_G$s, but there are large differences between 
the solid and dashed lines when $R_G=10h^{-1}{\rm Mpc}$.
However, we can make the corrections by multiplying 
a factor $((\sigma_0/{\sigma}_1)/(\hat\sigma_0/\hat{\sigma}_1))^i$
on $i$-th order MF $V_i$, and rescaling the density threshold from 
$\delta$ to $(\sigma_0/\hat{\sigma}_0)\delta$ to correct the shot 
noise effects on the measured MFs. We show the results after 
correcting as dotted lines, and find that they are well consistent
with the high-resolution results.
}

{
Similar to the statements in \cite{liu2023probing}, where they have tried 
to correct the shot noise effects on the MFs of the mock 
galaxy catalogue and find that the correction does not significantly
affect the Fisher forecast,
one should also not worry about how the shot noise will affect our 
Fisher results due to the same reason:
Shot noise effects mainly depend on the number density of 
the field and should play a similar role in different cosmology or 
gravity models, thus they almost cancel out during the process of 
evaluating the Fisher matrix. The density threshold $\delta$ 
is rescaled similarly for all the MFs present in both the 
derivative and covariance term of the Fisher matrix, and the amplitude 
change in the derivative term should be normalized by the inverse of 
the covariance matrix. Therefore, we below only show the results 
without the corrections.
}

\subsection{MG signatures in the MFs of halos}
\label{sec:halo mf}

\begin{figure*}
\centering
\includegraphics[width=1.0\textwidth]{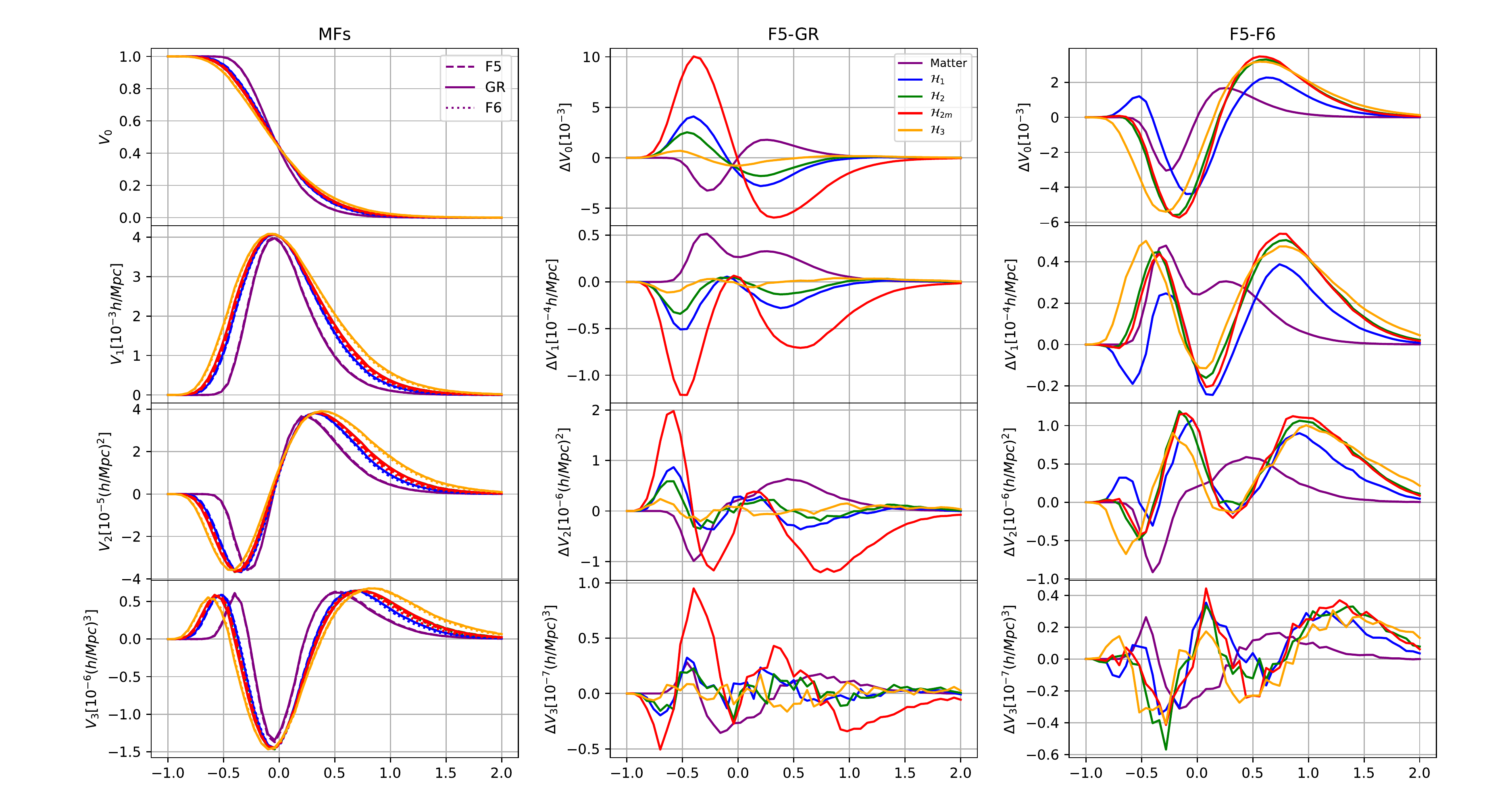}
\caption{\label{fig:MF_halo} {$Left$: The MFs measured 
from the Matter's and four halo populations' density fields
for GR, F6 and F5,
with $R_G=15h^{-1}\rm Mpc$ and at $z=0$. $Middle$: Differences in the MFs
between F5 and GR models. $Right$: Same as the middle panel but between
F5 and F6 models.}}
\end{figure*}

{We show $V_i$ and $\Delta{V_i}= V_i({\rm F5})-V_i({\rm GR})$ 
in the left and middle panels of figure~\ref{fig:MF_halo} respectively, 
which are measured from the distributions of the matter and
the four populations of halo at $z=0$.
The density field is smoothed with 
$R_G=15h^{-1}\rm Mpc$.
We do not show the error bars here and below
for which one can find the error level from the Fisher matrix
results. Additionally, we show the difference in the 
MFs between F5 and F6 in the right panel of figure~\ref{fig:MF_halo},
for an intuitive sense of the derivative term in the Fisher matrix.
Below we mainly focus on the left and middle panels to
understand the MG signatures.}

{We first discuss the overall influences of the halo bias,
that is the differences between the purple and other lines in the 
left pannel of figure~\ref{fig:MF_halo}.
One can directly find from the figure 13 in \cite{2021MNRAS.508.3771L}
the Gaussian assumption does not work for the biased field, 
where they show the MFs of the mock galaxies and the corresponding Gaussian field.
We refer the readers to \cite{2013MNRAS.435..531C,2000astro.ph..6269M} for detailed discussions of the 
bias effects on the MFs under a weakly non-Gaussian assumption
and to the section 4.3 in \cite{liu2023probing} for a brief summarization of these 
discussions.
More intuitively, the curves of the MFs of the matter field 
will expand a lot to both low and high ends 
of $\delta$ when transforming to the MFs of the halos due to 
a larger-than-one halo bias, and the larger the bias is, the wider they expand. 
Then the non-Gaussian properties,
for example, the asymmetry that two peaks in $V_3$ have different
widths and amplitudes, become more significant.
The different values of the zero-order spectral momentum
$\sigma_0$ for these fields 
will lead to the differences in the amplitudes of the last three MFs.
}

{
Next, we discuss the MG signatures encoded in the four MFs. We can
find from the middle panel that the four halo populations have
similar trend of $\Delta{V_i}$, but the trends are different from 
the matter's. 
We also find that the difference between 
F5 and GR becomes smaller as the halo number density decreases 
from $\mathcal H_1$ to $\mathcal H_3$. 
And the population with the same mass 
limit $\mathcal H_{2m}$ of F5 shows much 
larger differences from the GR case.  
To understand all these findings, we can look at a fixed mass 
limit of the halos, due to the enhancement of 
gravity there will be more halos originating from smaller density
peaks in $f(R)$ gravity when the initial conditions are the same for 
$f(R)$ and GR. Therefore, the overall model difference in halo clustering 
behaves differently from that for the matter. 
When the halo number density is fixed, 
theoretically, we are comparing the clustering properties of
the same sets of density peaks, thus the trend caused by the modified 
gravity will behave closely to the matter, as one can find 
from the right panel, where the behaviors are much similar for 
the halo and the matter field. In the next subsection, we'll explain the MG induced 
signatures implied in the four MFs one by one based on the 
$\Delta{V_i}$ (red) curves of $\mathcal H_{2m}$
}

\begin{figure*}
   \centering
   \includegraphics[width=1.0\textwidth]{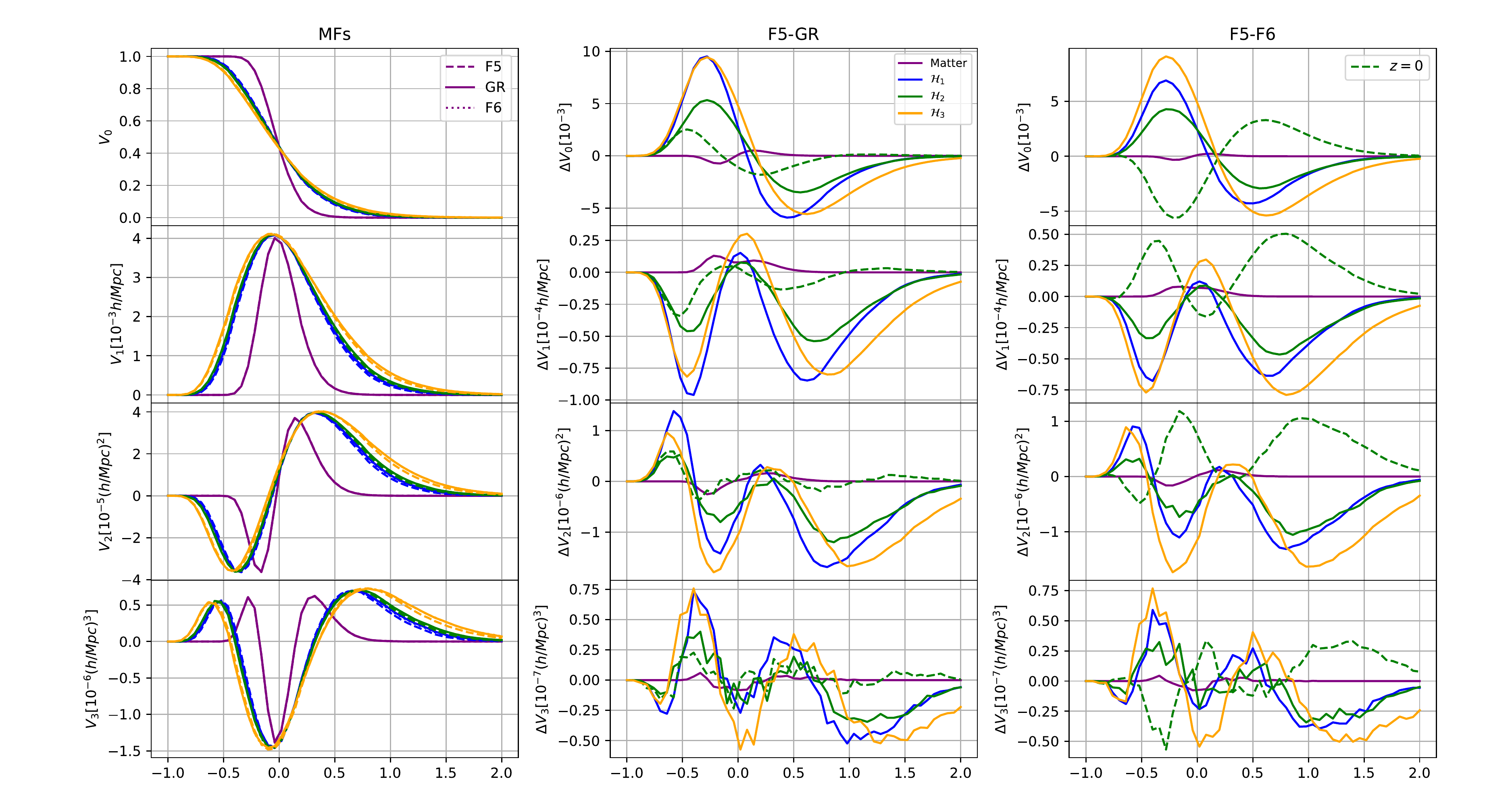}
   \caption{\label{fig:MF_halo_z1}{Same as 
   Figure.~\ref{fig:MF_halo} but at $z=1$. The green dashed lines in 
   the middle and right panels are the results for $\mathcal{H}_2$ at $z=0$ which we plot
   as a comparison.}
   }
\end{figure*}

{
We also show the results at $z=1$ in figure \ref{fig:MF_halo_z1}
to see how the MG signatures encoded in the MFs of halos change 
with redshifts. That is, the differences between 
F5 and GR detected from the halo
field at $z=1$ are more significant than those at $z=0$. 
Both higher and lower number density halo populations show 
more significant MG signatures than $\mathcal H_2$. As 
we have discussed above, these curves of $\Delta{V_i}$
receive great influence from the halo bias, e.g., the curves of 
$\Delta{V_0}$ actually indicate a smaller halo bias in $f(R)$
scenarios which we will discuss below. One can 
understand the results as the difference in halo bias between 
GR and $f(R)$ is mass and redshift dependent (see 
table~\ref{halo_population_properties}). In addition to 
the fact that the different halo populations may host different
type of galaxies, our results hint that properly choosing the 
type of galaxies in the future survey might provide us with 
better constraints on the $f(R)$ gravity.
}

{
We below focus on the results of $\mathcal H_{2m}$ 
in the figure \ref{fig:MF_halo} and compare them with those 
of the matter field, 
and clarify the morphological changes induced by the modified 
gravity.}

$V_0$ is the volume fraction occupied by the excursion sets. 
From the curve of $V_0$, we find that, compared to the matter 
field, for halos, regions with their number density above a 
threshold specified by a given $\delta$ occupy a larger 
fraction of total volume when $\delta\gsim0$. 
But they occupy a smaller volume fraction when $\delta\lsim0$, 
which indicates a larger volume fraction occupied by the 
under-dense regions. The results indicate an overall larger 
fluctuation for the halo number density field, 
which is consistent with the larger-than-one halo bias. 
From the curve of $\Delta{V_0}$, we find $\Delta{V_0}<0$ 
when $ \delta\gsim0$ and $\Delta{V_0}>0$ when $\delta\lsim0$ 
for halos. That is, the fractional volume occupied by regions 
with density above an overdense threshold becomes smaller 
in the MG scenario, while for regions above an under-dense 
threshold, it becomes larger. The latter indicates a smaller 
volume fraction occupied by regions with density below the 
underdense threshold. 
These results imply an overall smaller fluctuation 
for the halo field in the $f(R)$ gravity. 
To confirm the result, we measure the root mean square of 
the smoothed halo number density field {$R_G=15h^{-1}\rm Mpc$, 
and find it is $\sim 2.6\%$} smaller for the F5 model when 
compared with that for GR. 

However, the situation for the matter field is almost the 
opposite. As was discussed in \cite{2017PhRvL.118r1301F}, 
the curve of $\Delta{V_0}$ for matter indicates that halos 
and voids are larger and/or more abundant in the F5 model 
because of the enhancement of gravity. For halos, 
our results indicate relatively weak clustering properties, 
which is consistent with a relatively smaller halo bias 
in the F5 model \cite{2009PhRvD..79h3518S,2019MNRAS.483..790A}. 
As we state above, when the mass cut is the same, 
the halos in F5 form from smaller peaks in the initial 
density field, which are intrinsically less clustered. 
Another reason is that the bias is the ratio between halo 
clustering and matter clustering, and matter clustering is 
stronger in F5 than in GR. 
It is worth mentioning that the result for $\delta\lsim0$ 
indicates if voids are traced by halos, 
they might be smaller and/or less abundant in the MG scenario, 
which is consistent with \cite{2015MNRAS.451.1036C,2021MNRAS.504.5021C}. 

$V_1$ is the surface area of the excursion sets or the 
isodensity contours. The values of $V_1$ become smaller 
in the F5 model for nearly all density thresholds for the 
halo field. For a high enough threshold, one can assume that 
the excursion sets are isolated overdense regions enclosed by 
the isodensity contours. Hence it is natural to expect 
that $V_1$ becomes smaller given that the volume fraction 
occupied by these overdense regions becomes smaller. 
For a low enough density threshold, although one can still 
assume the isodensity contours are isolated, they no longer 
enclose the excursion sets but the regions with density below 
the threshold, whose volume fraction is $1-V_0$. 
Thus the surface area is expected to become smaller 
when $\Delta{V_0}>0$. The trend of $\Delta V_1$ is 
the opposite for the matter density field, but the logic is 
the same: the change of the surface area of the isodensity 
contours follows that of the volume fraction occupied by 
the regions they enclose.

{$V_2$ is the integrated mean curvature over the 
surface area of the excursion sets,
\begin{equation}
V_2(\delta)={1\over\mathcal{V}} \int v_2^{(\rm loc)}(\delta,\bm{x}) d^2S(\bm{x}),
\end{equation}
where $\mathcal{V}$ is the volume of the box, and $v_2^{\rm(loc)}(\delta,\bm{x})$ is the local mean curvature (average of the two principle curvatures) at a point $\bm{x}$ on the surface of the excursion sets with density threshold specified by $\delta$. From the definition, the MG effects on $V_2$ is determined by both its effects on the surface area, which we have already obtained, and on the mean curvature. As suggested in \cite{2022PhRvD.105j3028J,WeiLiu2021}, we can separate the two by rewriting $V_2$ as
\begin{equation}
V_2(\delta)=V_1(\delta)\left<v_2(\delta)\right>.
\end{equation}
Here the ratio of $V_2$ to $V_1$ gives the surface-area 
weighted average of the mean curvature 
$\left<v_2(\delta)\right>$.}
The sign of $V_2$ represents whether the surface is overall 
convex or concave, and its absolute value $|V_2|$ determines 
how much the surface is curved. 
Since we define the positive direction of the surface 
pointing from lower to higher density regions, 
$V_2$ is negative when $\delta\lsim-0.1$ and positive 
when $\delta\gsim-0.1$. By comparing the curves of $V_2$ for 
the two gravity models, we find $|V_2|$ is larger in the 
F5 model {when $-0.46\lsim\delta\lsim0.38$ for the halo field, 
and smaller elsewhere. }
The situation is different for the matter field, 
where $|V_2|$ is larger for nearly all density ranges 
in the MG scenario. 

\begin{figure}[ht!]
\centering
\includegraphics[width=1.\linewidth]{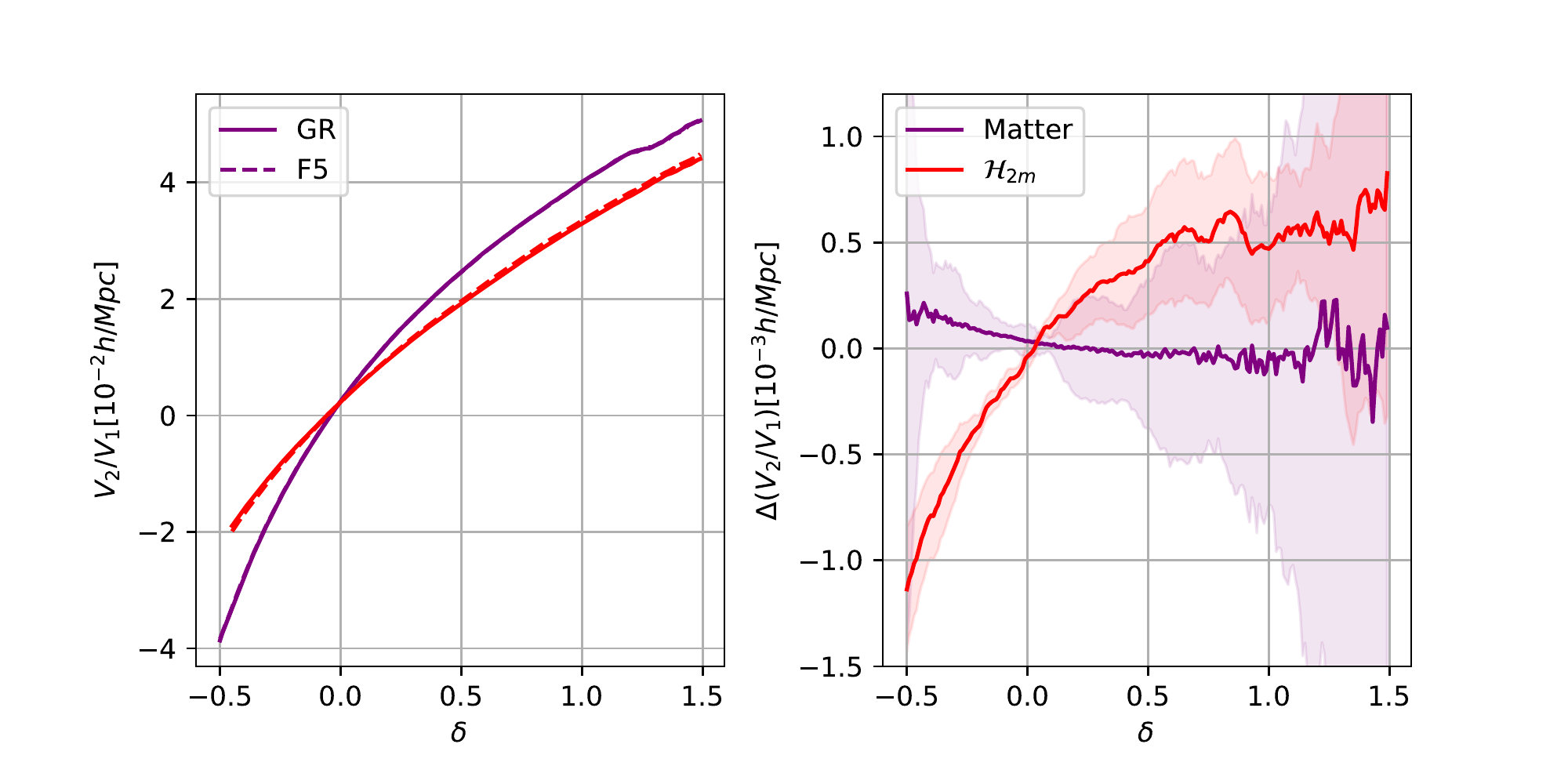}
\caption{\label{fig:ratio} {$Left$: 
Ratio of $V_2$ to $V_1$ for F5 (dashed) and GR (solid) 
measured from the halo (red, $\mathcal{H}_{2m}$) and 
matter (purple) density field with $R_G=15h^{-1}\rm Mpc$ and
at $z=0$, as a function of $\delta$. $Right$: 
The difference in this ratio between the two gravity models.}
}
\end{figure}

{To obtain the new information brought by $V_2$, 
we plot $V_2/V_1$ and the difference in it between 
the two gravity models in Fig.~\ref{fig:ratio}. 
The ratio is a global measure of the mean curvature of 
excursion sets.}
We take $\delta\lsim-0.5$, 
where both the values of $|V_2|$ and $V_1$ are smaller in 
F5 for halos, as an example to understand the curves. 
We can assume the under-dense regions are isolated and 
sphere-like structures in this density range, 
whose curvatures are inversely proportional to their radiuses. 
Compared to the GR case, the smaller $|V_2/V_1|$ in 
F5 for the matter field indicates the under-dense structures 
(e.g. voids traced by dark matter particles) on average have 
larger radiuses (in other words, voids are larger and/or 
large voids are more abundant). The larger $|V_2/V_1|$ for 
the halo field indicates the opposite results: 
voids traced by dark matter halos are statistically 
smaller in the F5 model, which is consistent with 
findings in \cite{2015MNRAS.451.1036C,2021MNRAS.504.5021C}.

The Euler characteristic $V_3$ describes the 
connectedness of the structures, 
which equals to the number of isolated structures 
minus the number of holes.  
We find for halos the excursion sets are more 
connected with $V_3<0$ when {$-0.34\lsim\delta\lsim0.32$ 
and less connected with positive $V_3$ elsewhere. 
From the curve of $\Delta{V_3}$, we find in the ranges 
$-0.52\lsim\delta\lsim-0.16$ and $0.08\lsim\delta\lsim0.7$, }
the structures are less connected in F5 with $\Delta{V_3}>0$, 
and more connected elsewhere.

\subsection{Constraint from the MFs of halos}
\label{sec:halo con}
To evaluate the constraining power from the MFs of the 
halo number density field, we also use the Fisher matrix 
to estimate the predictive error 
$\sigma_{{\rm log}_{10}(|f_{R0}|)}$. 
Similar to what we did for the matter density field, 
we divide the halo density field constructed from each 
original realization into 64 subsamples 
to estimate the covariance matrix. 
The constraints obtained from the halo field are also shown 
in Table.\ref{tab:constraint_value}. 

{We find that for all these halo populations, 
the variation of the constraint with $R_G$ is the 
same as that for the matter field. For each $R_G$,
the constraints show dependence on the redshifts and
halo populations. For example, the results obtained 
from $\mathcal H_2$ at $z=0$ are slightly better 
than the constraint from 
the same halo population at high redshift. But the results 
from the halo population with the lowest number density 
provide us with the best constraint at $z=1$. These findings 
are consistent with the results shown in Fig.~\ref{fig:MF_halo} 
and \ref{fig:MF_halo_z1}.}

Similar forecasts for the constraints on the $f(R)$ modified gravity parameters have been done in the literature by using more traditional statistics, e.g., using the
  galaxy clustering and cluster abundance \cite{2021PhRvD.104j3519L}, galaxy clustering and weak lensing \cite{2019PhRvD.100l3540W}, and cosmic shear statistics \cite{2017MNRAS.466.2402S, bose2020road}. Specifically, \cite{2021PhRvD.104j3519L} conducts a Fisher analysis using $\sigma_8$ constraints derived from the abundance of thermal Sunyaev-Zel'dovich effect selected galaxy clusters from the Simons Observatory as well as linear and quasilinear redshift-space 2-point galaxy correlation functions from a DESI-like experiment. For a log-based fiducial parameter value of $\log_{10}(|f_{R0}|)=-5$, paired with the parameter value $n=1$, they find a predicted error $\sigma_{{\rm log}_{10}(|f_{R0}|)}=0.12$ and $\sigma(n)=0.36$ after combining the galaxy clustering and cluster abundance (see Table II in \cite{2021PhRvD.104j3519L}). Ref \cite{2019PhRvD.100l3540W} is even a step forward attempt, which tries to establish a nonlinear matter power spectrum emulator for modified gravity models based on N-body simulations.
For a Euclid-like mission and fiducial value of $\log_{10}(|f_{R0}|)=-5$, they obtain a constraint of $\sigma_{\log_{10}(|f_{R0}|)}=0.2$ by using the redshift-space galaxy power spectrum with $k_{max}=0.25h/\rm Mpc$, which can be tightened to $\sigma_{\log_{10}(|f_{R0}|)}=0.04$ when combined with Euclid weak lensing (see Table III in \cite{2019PhRvD.100l3540W}).

{Both these results look better than ours. However, the survey volume we forecast for is much smaller than theirs, which is $1.56\times10^{-2} h^{-3}\rm Gpc^3$, while that for Euclid is $\sim68h^{-3}\rm Gpc^3$, for DESI is $\sim61h^{-3}\rm Gpc^3$. According to our findings in Sec.\ref{sec:volume}, the constraints we obtain will be improved by approximately a factor of 66 or 63, reaching the accuracy level of $10^{-3}$, should we assume a survey volume of Euclid or DESI. Considering we have only 1 parameter in our forecast, these constraints will be degraded to some degree in a multi-parameter forecast. However, our findings still highlight the strong constraining power of the MFs on MG parameters compared to the traditional statistics. }

\section{Redshift space distortion and MFs of HOD Galaxies}
\label{sec:rsd+HOD}

{
In this section, we discuss how the redshift space distortion (RSD) in 
Sec.~\ref{sec:rsd},
which is one of the most important effects for any cosmological 
analysis from the spectroscopic galaxy surveys. And based on 
the mocked galaxy catalogue using the halo occupation distribution 
method, we briefly study the modified gravity signatures detected 
by the MFs of the galaxy distributions.
}

\subsection*{Redshift space distortion}
\label{sec:rsd}

\begin{figure*}
   \centering
   \includegraphics[width=1.0\textwidth]{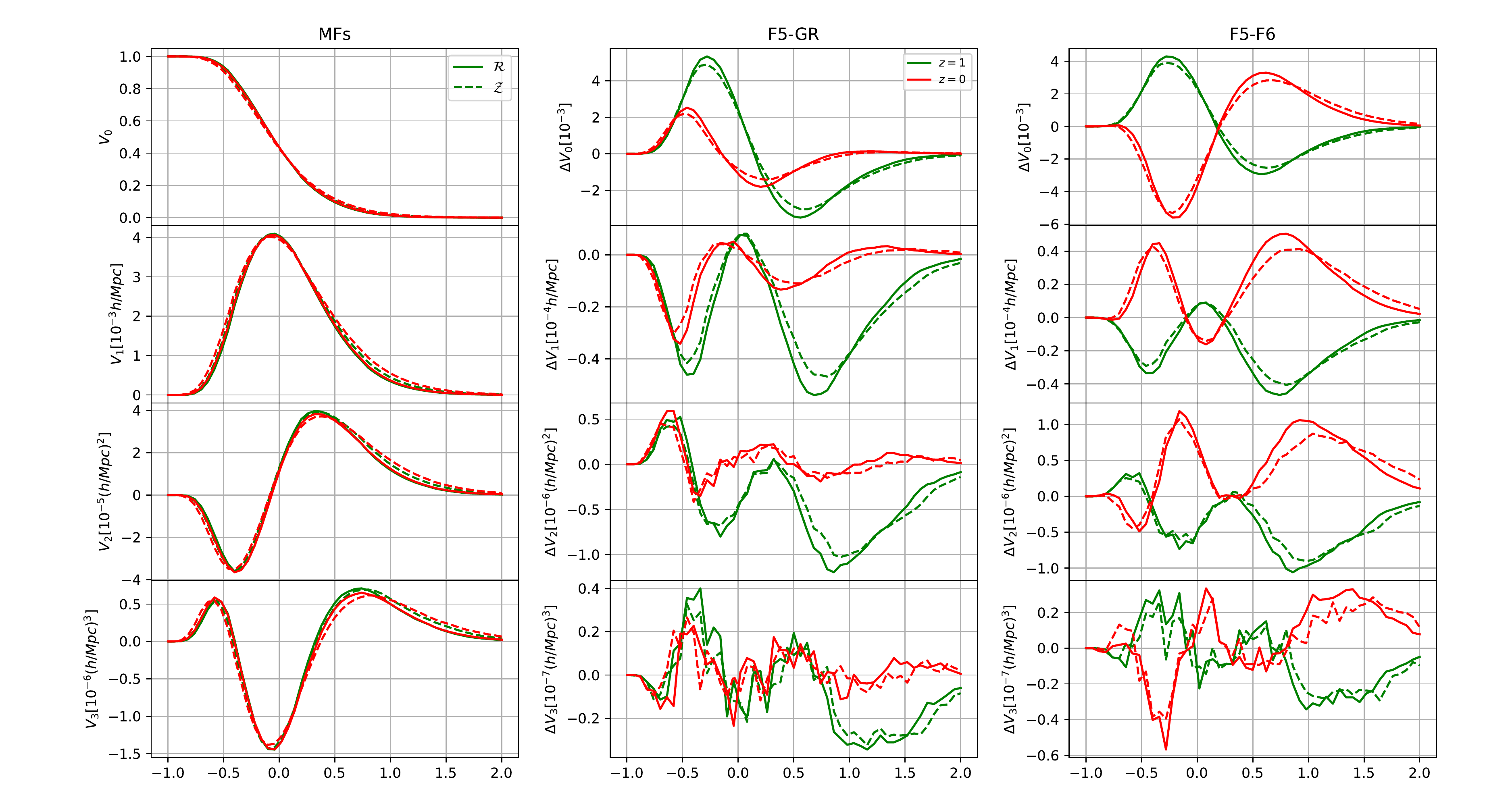}
   \caption{\label{fig:rsd} {$Left:$ The MFs measured 
   from $\mathcal{H}_2$ of GR model in real ($solid$) and redshift
   ($dashed$) space and at $z=0$ ($red$) and $z=1$ ($green$). 
   $Middle$ and $right$: Similar to the panels in 
   Figure.\ref{fig:MF_halo}.}}
\end{figure*}

{
The two-point redshift space clustering in modified gravity
scenarios have been widely studied in \cite{2015PhRvD..91f3008X,
2012MNRAS.425.2128J,2019MNRAS.485.2194H,2013MNRAS.436...89R,
2019NatAs...3..945A, 2017PhRvD..96b3519B,2016JCAP...08..032B}. For example,
works in \cite{2019MNRAS.485.2194H} find that the linear 
model cannot model the RSD well in both GR and modified gravity scenarios 
on small scales, and the linear distortion parameter 
$\beta$ can hardly be used to limit the $f(R)$ gravity.
But in \cite{2021PhRvD.103j3524G} they find if a proper estimator 
of $\beta$ is chosen, a significant difference
between MG and GR models can be extracted.
}

{
The effects of redshift space distortion on the MFs under the
Gaussian limit is studied in \cite{1996ApJ...457...13M} 
where they find by choosing
a unique threshold $\nu_A$ to define the patterns, the MFs in redshift
space can have a similar shape but different amplitude compared 
to those in real space, while the non-Gaussian case is also
detailedly studied in \cite{2013MNRAS.435..531C}. As one of the most import 
effects when applying the MFs to limit the MG parameters in 
the galaxy survey, we also study how the RSD effects affect
our measured MFs.
}

{
We use a simple distant-observer approximation and choose 
the $z-$axis as the line of sight to 
get the position $\vec{s}$ of 
the halos/galaxies in redshift space,
\begin{equation}
   \vec{s}=\vec{r}+(1+z)\vec{v}_{\parallel}/H(z),
\end{equation}
where $\vec{r}$ and $\vec{v}_\parallel$ are their real
space position and the line-of-sight component of their 
peculiar velocities, while $H(z)$ is the Hubble parameter 
at redshift $z$. We choose $\mathcal{H}_2$ as a representative
and show its MFs at the two redshifts in figure \ref{fig:rsd}
}

{
We find that the RSD effects not only reduce the amplitude of the MFs
with orders 1, 2, and 3, but also expand the curves of MFs from 
$\delta\sim1$ toward both higher and lower ends of thresholds.
The reason for the former is that the MFs in redshift space
are sensitive to the parameter $fb^{-1}$ \cite{2013MNRAS.435..531C} where $f$ is the growth
rate and $b$ is the bias. The latter is because the RSD enlarges
the variance of the field, and the MFs as a function of $\delta$
become wider as the variance of the field gets larger.
}

{
From the middle and right panels in figure \ref{fig:rsd}, 
we find that the RSD effects slightly reduced the differences
between models. The reason is that the MG-induced clustering
properties are degenerated to the RSD effects: The clustering in real space
is weaker in $f(R)$ but the RSD effects which enlarges the clustering 
is relatively stronger. We also perform
a Fisher matrix analysis to quantify the detected MG signatures
in redshift space and show the results in table \ref{tab:constraint_value}.
The findings are consistent in that the RSD effects slightly reduce the
constraining power of the MFs.
}

\subsection*{HOD galaxies}
\label{sec:HOD}

\begin{figure*}
   \centering
   \includegraphics[width=1.0\textwidth]{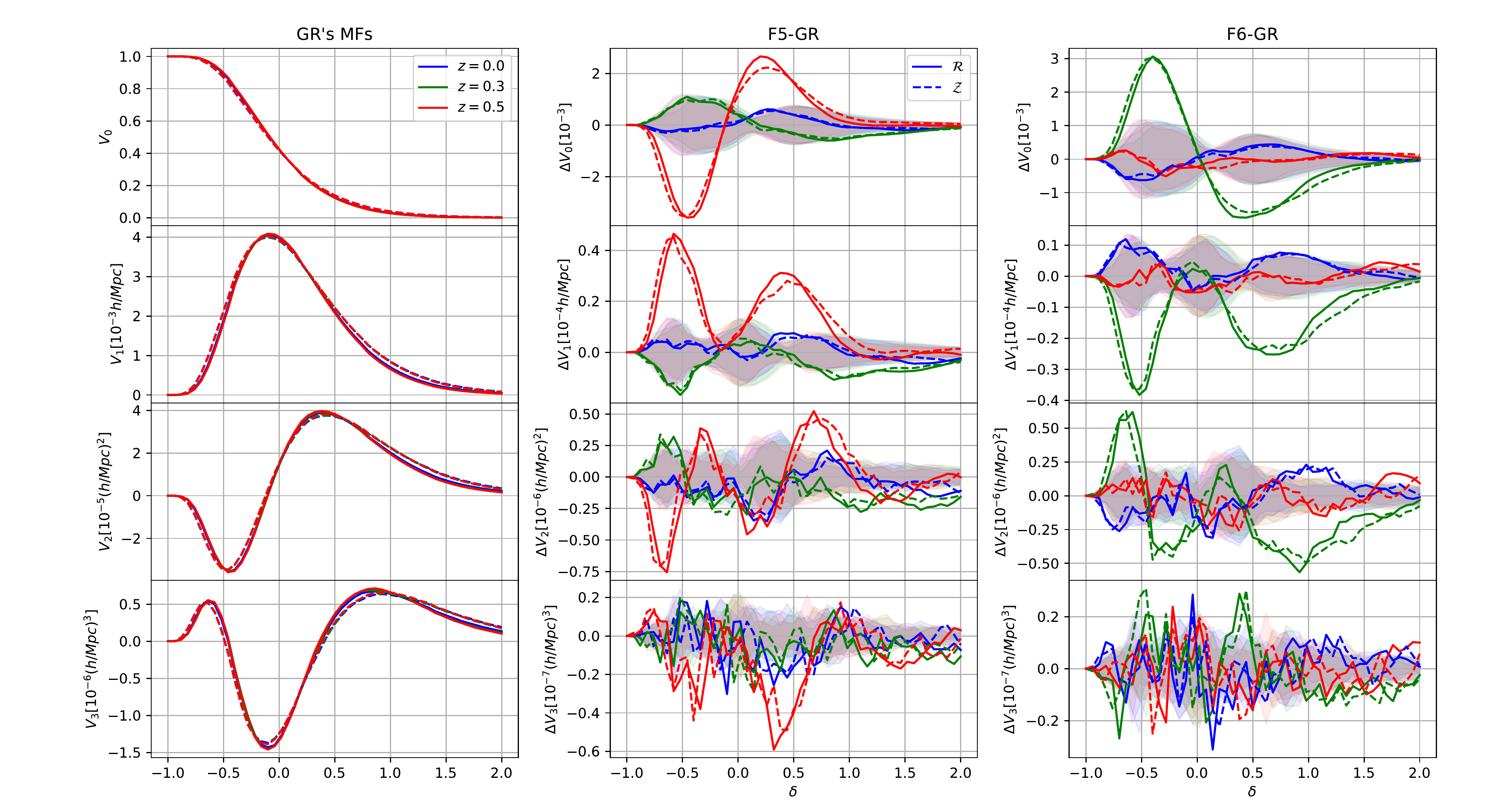}
   \caption{\label{fig:hod} {$Left$: MFs 
   measured from HOD galaxies of GR model with $R_G=15h^{-1}\rm Mpc$
   in real ($solid$) and redshift ($dashed$) space.
   $Middle$/$Right$: The difference in MFs between F5/F6 and 
   GR models.}
   }
\end{figure*}
{
We build the galaxy catalogue using the halo population distribution (HOD)
models \cite{2000MNRAS.318.1144P,2002ApJ...575..587B,2007ApJ...667..760Z} with five parameters as \cite{2007ApJ...667..760Z}. In 
this model, the mean number of galaxies in a halo with mass
$M_h$ is the sum of the mean number of central and satellite
galaxies,
\begin{equation}
<N(M_h)>=<N_c>+<N_s>,
\end{equation}
the mean central galaxy occupation is given by,
\begin{equation}
<N_c(M_h)>={1\over2}\left[ 1+
{\rm erf}({\log M_h -\log M_{min} \over \sigma_{\log M}})\right]
\end{equation}
while the mean number of satellite galaxy is,
\begin{equation}
   <N_s(M_h)>=<N_c(M_h)>\times\left({M_h-M_0\over M_1}\right)^{\alpha},
\end{equation}
and $<N_s>=0$ if $M_h<M_0$. The central galaxies are placed
at the centre of their host halos while the satellite galaxies
are assumed radially distributed within radius $r_{200c}$ 
following the NFW profiles of their host halos.
}

{
   The five independent parameters \{$M_{min}$,$M_0$,$M_1$
   $\sigma_{\log M}$,$\alpha$\} for GR are chosen from
   the BOSS-CMASS-DR9 sample at $z=0.5$ \cite{2012MNRAS.427.3435A}. The parameters 
   for $f(R)$ models are tuned to match the number density and
   real-space clustering in the GR case. We refer who interests 
   to \cite{2019MNRAS.485.2194H} for the values of these 
   parameters and the detailed mock procedures. 
   We show the MFs measured from the mock galaxies 
   with $R_G=15h^{-1}\rm Mpc$ at three redshifts 0,0.3 and 0.5
   in figure \ref{fig:hod}.   
}

{
Given the process of tuning the HOD parameters to align with
the identical real-space clustering properties of GR and 
MG models, together with the 
degeneracy mentioned above, we may expect no MG signature
in the two-point statistics measured from the HOD galaxies. 
But as is shown in 
figure \ref{fig:hod}, where we show the errors measured from 
the fluctuations of the five \texttt{ELEPHANT} HOD mocks,
there are significant differences 
in the MFs between MG and GR models at some specific redshifts,
e.g., $z=0.5$ for F5 and $z=0.3$ for F6. 
The finding proves the MFs' power to 
capture the higher-order information and their potential 
to be used as a probe of modified gravity in the future 
galaxy surveys.
}

\section{Conclusion}
\label{sec:conclu}


{
Examing models of modified gravity aimed at addressing the cosmic acceleration 
problem is crucial for advancing our understanding of both gravity and cosmic 
acceleration. These theories imprint distinct signatures on the large-scale 
structure of the universe. In an effort to fully capitalize on upcoming precise 
large-scale structure surveys and derive stringent constraints on these theories,
we explore the potential of using the Minkowski functionals of the large-scale 
structure as a novel probe for modified gravity. Employing N-body simulations,
we investigate how the signal of modified gravity in the MFs of the large-scale 
structures is affected by smoothing scale, redshift, tracer bias and reshift space 
distortion. To quantify the constraining power, we use the Fisher matrix exploring 
how the constraints on modified gravity parameters change with various factors.
Our findings, summerized below,
}


\begin{itemize}
\item {By varying the survey volume V, 
we explicitly check how the constraint changes, 
and find it roughly scales with V by 
$\sigma_{{\rm log}_{10}(|f_{R0}|)}\propto 1 / \sqrt{V}$.}
\item {We analyze how the MG signals in the MFs of LSS change with 
smoothing scale and redshift, and find the signals are stronger for 
smaller smoothing scale and lower redshift. At the same time, 
we find that the forecasted error monotonically increases with $R_G$ at 
a given redshift, and for a given $R_G$, the constraint is better at a 
lower redshift.} 
\item {Because the MG effects in the MFs of LSS are more pronounced on small scales 
and at low redshifts, the combined constraint from different smoothing scales 
and redshifts is mostly dominated by the result from small smoothing scale and 
low redshift. 
This suggests that we should try to push to small scales and focus on low 
redshifts in future analysis.}
\item {To study how the tracer bias affects the results, we 
construct halo catalogues with different number densities at two redshifts. We find that 
the MG signatures are strongly affected by the mass and redshift denpendent halo 
bias.}
\item {We also study how the redshift space 
distortion affects our results. Due to the degeneracy between 
the bias and the redshift space distortion, the RSD effects will
slightly reduced the MG signatures encoded in the MFs.}
\item {Although the HOD parameters are tuned
to match the real space clutering properties between GR and $f(R)$,
we still find significant MG signatures contained in the MFs of the 
HOD galaxies, which indicates the MFs' power to extract high-order 
information.}

\end{itemize}

{Note that the constraints in this work are 
forecasted for a survey volume of 
$V\simeq1.56\times10^{-2} h^{-3}\rm Gpc^3$, 
which is much smaller compared to the next-generation galaxy 
redshift surveys. We expect much better constraints with a 
volume as large as the next-generation surveys. 
Our best constraint from an individual scale is obtained 
for $R_G=5h^{-1}\rm Mpc$, which is completely accessible 
for the currently on-going DESI survey. 
Our analysis in this work are all based on N-body simulations. 
Systematical effects such as irregular survey geomtry, 
which will affect our results, 
have not been taken into account.  
Although our studies focus on the MG parameter, 
they can be instructive for the general 
application of MFs as a cosmological probe.
}

\begin{acknowledgments}

This work is supported by the National Natural Science Foundation of China Grants No. 12173036 and 11773024, by the National Key R\&D Program of China Grant No. 2021YFC2203100 and No. 2022YFF0503404, by the China Manned Space Project “Probing dark energy, modified gravity and cosmic structure formation by CSST multi-cosmological measurements” and Grant No. CMS-CSST-2021-B01, by the Fundamental Research Funds for Central Universities Grants No. WK3440000004 and WK3440000005, by Cyrus Chun Ying Tang Foundations, and by the 111 Project for "Observational and Theoretical Research on Dark Matter and Dark Energy" (B23042).
BL is supported by the European Research Council (ERC) through a starting Grant (ERC-StG-716532 PUNCA), and the UK Science and Technology Facilities Council (STFC) Consolidated Grant No. ST/I00162X/1 and ST/P000541/1.
CB-H is supported by the Chilean National Agency of Research and Development (ANID) through grant FONDECYT/Postdoctorado No. 3230512.

This work used the DiRAC@Durham facility managed by the Institute for Computational Cosmology on behalf of the STFC DiRAC HPC Facility (www.dirac.ac.uk). The equipment was funded by BEIS via STFC capital grants ST/K00042X/1, ST/P002293/1,
ST/R002371/1 and ST/S002502/1, Durham University and STFC operation grant ST/R000832/1. DiRAC is part of the UK National e-Infrastructure.

\end{acknowledgments}

\begin{appendix}



  \section{Convergence test}
  \label{append:covariance}
  
\begin{figure}[htbp] 
     \begin{center}
     \includegraphics[width=1.0\linewidth, angle=0]{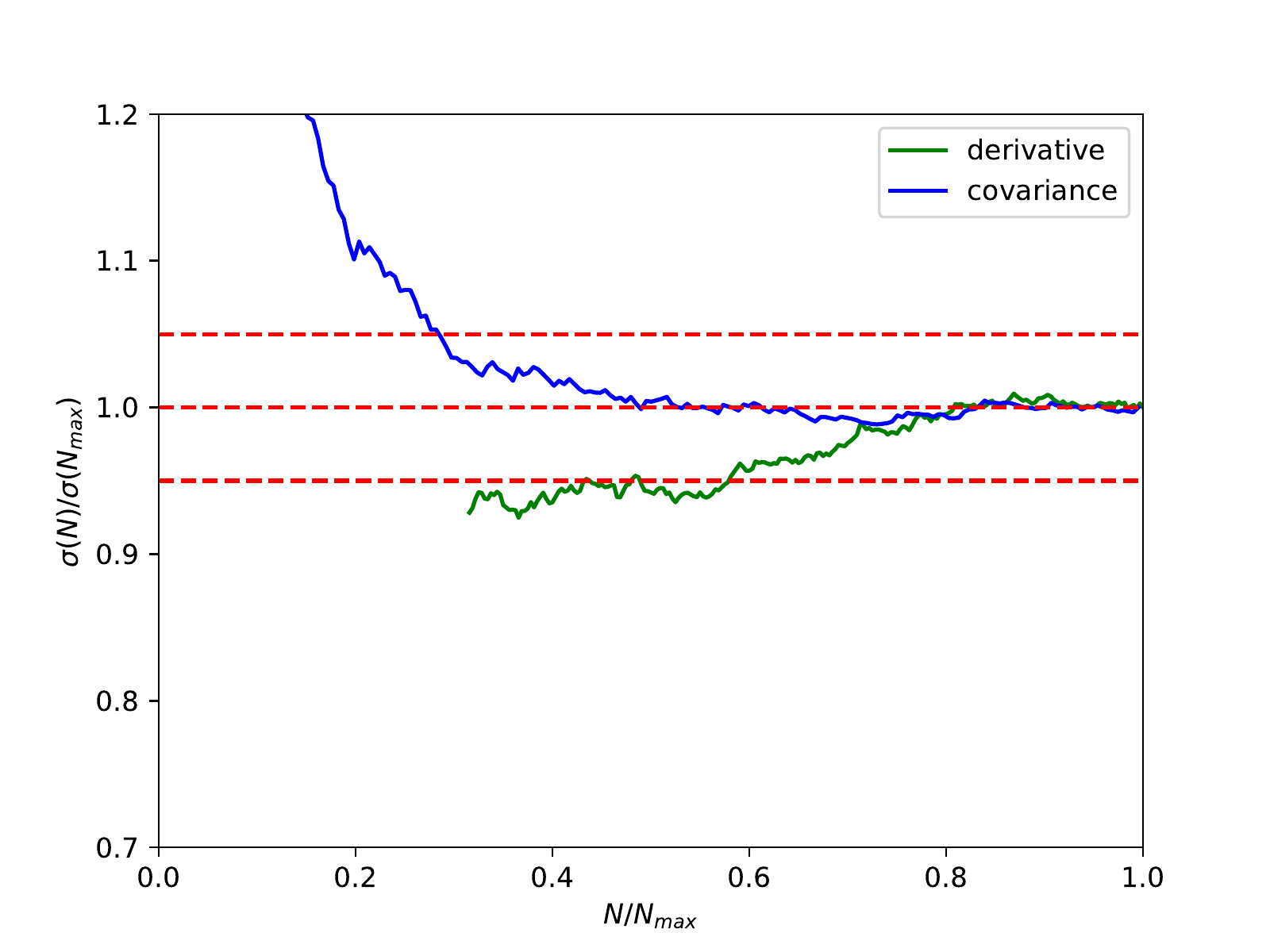} 
     \caption{{Convergence of the error $\sigma_{{\rm log}_{10}(|f_{R0}|)}$ from the MFs ($R_G=5h^{-1}\rm Mpc$). The blue line shows the ratio of the constraint obtained from the covariance matrix estimated from $N$ sub-boxes to that obtained from $N_{max}=1920$ sub-boxes, while the number of sub-boxes to estimate the derivative term is fixed. The green line shows the ratio of the constraint obtained with the derivative estimated from $N$ sub-boxes to that obtained from $N_{max}=320$ sub-boxes, while the number of sub-boxes to estimate the covariance term is fixed.} }
     \label{fig:cov_valid}
     \end{center}
  \end{figure}

 One may be concerned about whether the Fisher matrix converges because of our limited number of samples. As shown in Eq. 3.1, the Fisher matrix contains two terms: the derivative term and the covariance term. Hence we do the convergence test separately. For the covariance term, we fix the number of the subsamples for derivative estimation and each time select $N$ subsamples from $N_{max}=1920$ sub-boxes for covariance estimation.
For the derivative term, we fix the number of the subsamples for covariance estimation and randomly select $N$ subsamples from $N_{max}=320$ sub-boxes each time without replacement, then estimate the derivative term from them. We obtain these subsamples by dividing each ELEPHANT GR realization into 64 sub-boxes with $L_{box}=256h^{-1}\rm Mpc$. We note due to the additivity, there should be little difference between the mean values of the MFs estimated from the original 5 GR realizations and their subsamples.
We show in Fig.\ref{fig:cov_valid} how the errors from the MFs converge when the number of samples is increased. We find that when $N_{covariance}>300$ and $N_{derivative}>200$, the error converges (with $|\sigma(N)/\sigma(N_{max})-1.0|<0.05$). 
    
 \section{covariance dependency on the parameters}
 \label{covariance_parameters}

\begin{table}[ht!]
 \centering
  \begin{tabular}{c|cccc}   
  \toprule
  \hline
 $R_G[h^{-1}{\rm Mpc}]$ &5  & 10 &15 & 20\\
 
\hline
\texttt{RAMSES},{GR} &0.28&0.73&1.69&2.96\\
\hline
\texttt{ELEPHANT}, {GR} & 0.27 & 0.70 & 1.52 & 2.56\\
 \hline
 \texttt{ELEPHANT}, {F5}  & 0.27 & 0.68 & 1.54 &2.59\\ 
\hline
\bottomrule
     \end{tabular}
\caption{ {
To test our assumption that the covariance matrix is independent
with the cosmological and modified-gravity parameters, we 
divide the five \texttt{ELEPHANT} realizations into 320 sub-boxes 
and obtained the Fisher matrix results from them. The table shows
values of $\sigma_{\log(|f_{R0}|)}$ in the two cases.
}}
 \label{tab:2}

\end{table}

{
It is a common choice assuming that the covariance matrix of the observables 
is independent with the cosmological and modified gravity parameters.
However, since we choose F5 as our fiducial model but estimate the covariance 
using the GR realisations, concerns 
are raised regarding its validity of such choice.
In order to address these concerns, we partitioning the 5
\texttt{ELEPHANT} GR realizations into 320 sub-boxes and 
applying the same procedure to the 5 F5 realizations. 
Subsequently, we estimate the covariance matrix at $z=0$ 
for each $R_G$ from the respective subsamples.
}
{
The outcomes of this analysis are presented in the table \ref{tab:2},
and we also show the related values obtained from the \texttt{RAMSES} realizations in the table.
The differences caused by the different cosmological or 
modified gravity parameters are minor, which proves the 
validity of our choice.
}

\end{appendix}

\bibliographystyle{JHEP}
\bibliography{reference.bib}

\end{document}